\documentclass[sigconf, nonacm]{acmart}

\newcommand\vldbdoi{XX.XX/XXX.XX}
\newcommand\vldbpages{XXX-XXX}
\newcommand\vldbvolume{18}
\newcommand\vldbissue{1}
\newcommand\vldbyear{2025}
\newcommand\vldbauthors{\authors}
\newcommand\vldbtitle{\shorttitle} 
\newcommand\vldbavailabilityurl{https://www.doi.org/10.6084/m9.figshare.28518449}
\newcommand\vldbpagestyle{plain}

\newcommand*\circled[1]{\tikz[baseline=(char.base)]{\node[shape=circle,draw,inner sep=0.85pt, fill=black] (char) {\textcolor{white}{#1}};}}

\newcommand{\textt}[1]{{\fontfamily{lmss}\selectfont #1}}

\usepackage{tabularx}
\usepackage[noend]{algpseudocode}
\usepackage{algorithm}
\usepackage{tcolorbox}
\usepackage{tikz}
\usepackage{amsfonts}
\usepackage{eucal}
\usepackage{graphicx}

\usepackage{graphicx}
\usepackage{subcaption}

\usepackage{algorithmicx}
\usepackage{underscore}

\usepackage{enumitem}

\usepackage{listings}
\usepackage[english]{babel}
\definecolor{codegreen}{rgb}{0,0.6,0}
\definecolor{codegray}{rgb}{0.5,0.5,0.5}
\definecolor{codepurple}{rgb}{0.58,0,0.82}
\definecolor{backcolour}{rgb}{0.95,0.95,0.92}
\lstdefinestyle{mystyle}{
    backgroundcolor=\color{backcolour},
    commentstyle=\color{codegreen},
    keywordstyle=\color{magenta},
    numberstyle=\tiny\color{codegray},
    stringstyle=\color{codepurple},
    basicstyle=\ttfamily\footnotesize,
    breakatwhitespace=false,
    breaklines=true,
    captionpos=b,
    keepspaces=true,
    numbers=left,
    numbersep=5pt,
    showspaces=false,
    showstringspaces=false,
    showtabs=false,
    tabsize=2
}

\newcommand*{\img}[1]{%
    \raisebox{-.3\baselineskip}{%
        \includegraphics[
        height=\baselineskip,
        width=\baselineskip,
        keepaspectratio,
        ]{#1}%
    }%
}

\MakeRobust{\Call} 

\def\name{\textt{Mayura}}

\newcommand{\FuncDef}[1]{\textcolor{magenta}{\textbf{#1}}} 
\newcommand{\FuncCall}[1]{\textcolor{magenta}{#1}} 
\newcommand{\Comm}[1]{\Comment{\textcolor{teal}{#1}}}

\begin{document}
\title{\name: Exploiting Similarities in Motifs for Temporal Co-Mining}

%
\author{Sanjay Sri Vallabh Singapuram}
\affiliation{%
  \institution{University of Michigan}
  \city{Ann Arbor}
  \state{Michigan, USA}
}
\email{singam@umich.edu}

\author{Ronald Dreslinski}
\affiliation{%
  \institution{University of Michigan}
  \city{Ann Arbor}
  \state{Michigan, USA}
}
\email{rdreslin@umich.edu}

\author{Nishil Talati}
\affiliation{%
  \institution{University of Michigan}
  \city{Ann Arbor}
  \state{Michigan, USA}
}
\email{talatin@umich.edu}

%
\begin{abstract}
Temporal graphs serve as a critical foundation for modeling evolving interactions in domains ranging from financial networks to social media.
Mining temporal motifs is essential for applications such as fraud detection, cybersecurity, and dynamic network analysis.
However, conventional motif mining approaches treat each query independently, incurring significant redundant computations when similar substructures exist across multiple motifs.
In this paper, we propose \name, a novel framework that unifies the mining of multiple temporal motifs by exploiting their inherent structural and temporal commonalities.
Central to our approach is the \emph{Motif-Group Tree (MG-Tree)}, a hierarchical data structure that organizes related motifs and enables the reuse of common search paths, thereby reducing redundant computation.
We propose a co-mining algorithm that leverages the MG-Tree and develop a flexible runtime capable of exploiting both CPU and GPU architectures for scalable performance.
Empirical evaluations on diverse real-world datasets demonstrate that \name~achieves substantial improvements over the state-of-the-art techniques that mine each motif individually, with an average speed-up of 2.4$\times$ on the CPU and 1.7$\times$ on the GPU, while maintaining the exactness required for high-stakes applications.
\end{abstract}


\maketitle

\pagestyle{\vldbpagestyle}
\begingroup\small\noindent\raggedright\textbf{PVLDB Reference Format:}\\
\vldbauthors. \vldbtitle. PVLDB, \vldbvolume(\vldbissue): \vldbpages, \vldbyear.\\
\href{https://doi.org/\vldbdoi}{doi:\vldbdoi}
\endgroup
\begingroup
\renewcommand\thefootnote{}\footnote{\noindent
This work is licensed under the Creative Commons BY-NC-ND 4.0 International License. Visit \url{https://creativecommons.org/licenses/by-nc-nd/4.0/} to view a copy of this license. For any use beyond those covered by this license, obtain permission by emailing \href{mailto:info@vldb.org}{info@vldb.org}. Copyright is held by the owner/author(s). Publication rights licensed to the VLDB Endowment. \\
\raggedright Proceedings of the VLDB Endowment, Vol. \vldbvolume, No. \vldbissue\ %
ISSN 2150-8097. \\
\href{https://doi.org/\vldbdoi}{doi:\vldbdoi} \\
}\addtocounter{footnote}{-1}\endgroup


\section{Introduction}

Temporal graphs have become a fundamental abstraction for modeling dynamic interactions in domains ranging from financial transaction networks to social media ecosystems~\cite{Sommers2024,Fisher2024,TigerGraph2022,Smith2023,BalanRege2017,ZACHLOD20221064}.
These graphs capture not only topological relationships but also the temporal evolution of interactions, enabling the analysis of complex phenomena such as information diffusion, fraud patterns, and network dynamics.
With the advent of large-scale temporal datasets—exceeding billions of edges in domains like blockchain transactions and communication networks—the need for efficient analytical primitives has never been more critical~\cite{hu2021ogb}.
Traditional graph mining techniques, designed for static graphs, fail to account for the temporal ordering and time-window constraints inherent in real-world systems, necessitating specialized approaches for temporal pattern discovery~\cite{arabesque,peregrine,peregrineFollowup,graphzero,automine,graphpi,stmatch22, G2Miner2022OSDI,rapidmatch}.

A cornerstone of analyzing relationships in temporal graph is temporal motif mining, which identifies and enumerates sequences of time-constrained edges whose structure is governed by a motif (\textit{e.g.,} 3-cycles)~\cite{paranjape2017motifs}.
These motifs can represent meaningful relationships in the underlying temporal graph, such as a pattern of suspicious transactions between bank accounts within a short period of time, thus helping with fraud detection in financial networks~\cite{hajdu2020temporal}.
Applications of temporal mining also include cybersecurity threat analysis~\cite{mackey2018chronological,glasser2013bridging}, social behavior modeling~\cite{BalanRege2017,ZACHLOD20221064}, and monitoring energy disaggregation in electrical grids~\cite{shao2013temporal}.
Existing systems focus on mining individual motifs, while real-world workloads often need to process queries with multiple motifs that overlap structurally and temporally.
For instance, anti-money laundering investigations~\cite{blanuvsa2024graph,altman2023realistic,Sommers2024,Fisher2024} often require simultaneous detection of multiple transaction patterns across shared subsets of edges.
Current approaches thus incur redundant computations as they need to repeatedly traverse the graph for each motif.

While multi-query optimization (MQO) techniques for static graphs and approximate temporal mining offer partial solutions, they prove inadequate because the techniques used to exploit similarities are not applicable to temporal motif mining~\cite{ren2016multi,jamshidi2023accelerating}.
Approximate counting~\cite{teacups,edgesampling,sarpe2021presto,liu2019sampling,sarpe2021oden} sacrifices accuracy for performance, which cannot be used in high-stakes domains like finance, where the exact identification of crime is necessary.

This paper addresses aforementioned limitations by proposing \textbf{\name}, the first system to enable efficient co-mining of temporal motifs.
\name~accomplishes this by using a data-structure called the \textit{Motif-Group Tree (MG-Tree)}, a hierarchical representation of motifs that captures edge-level commonalities, enabling shared search path exploration and eliminating redundant computations.
The temporal mining algorithm is then adapted to search for matches guided by the MG-Tree instead of a single motif.
\name~supports efficient co-mining on both CPUs and GPUs, balancing the workload across multiple threads (and warps) and exploiting the hierarchical parallelism exposed by the MG-Tree.
\name~also generates code that optimizes execution for a specific MG-tree to improve instruction throughput and reduce the architectural resource footprint.

Co-mining presents several critical system-level challenges that must be addressed to maximize the efficacy of this approach.
Load balancing emerges as a fundamental challenge due to the inherent irregularity of graph workloads, where different search paths exhibit vastly different computational requirements, necessitating sophisticated work distribution strategies.
Co-mining on the GPU presents additional challenges.
The irregularity of the graph workload makes the GPU implementation susceptible to control-flow divergence, leading to serialized execution and lower performance, requiring the control-flow to be more streamlined than the baseline.
The GPU also poses a resource bottleneck with limited register capacity per thread when maintaining context for multiple motifs simultaneously, leading to reduced occupancy and suboptimal hardware utilization, and must be addressed by careful selection of instructions to minimize register usage.

Our comprehensive evaluation demonstrates \name's significant performance improvements across diverse temporal datasets.
On a 40-core Intel Xeon CPU,~\name~achieves average speedups between 1.8-3.7$\times$ over multiple datasets, with peak acceleration of 8.8$\times$ over a bipartite graph.
On an NVIDIA A40 GPU,~\name~achieves average speedups between 1.3-3.1$\times$ over many datasets, and 7.6$\times$ maximum speedup despite architectural constraints.
These gains stem from \name's core contributions: dynamic instruction counts reduce by 1.6-4.5$\times$ through the MG-Tree-guided search space pruning and code optimization, while motif-specific code generation alleviates 87-94\% of warp divergence penalties on the GPU.
These results validate that strategic co-mining of temporal motifs enables significant efficiency gains while preserving exactness for critical applications.
\name~makes the following contributions.
\begin{enumerate}
    \item  \textbf{The MG-Tree}: A hierarchical data-structure that captures structural and temporal similarities across motifs, helping unearth shared search paths.
    \item \textbf{Temporal Co-Mining Algorithm}: The first exact algorithm capable of simultaneously mining multiple motifs.
    \item \textbf{Multi-Backend Runtime}: A unified execution framework supporting both CPU and GPU backends.
    \item \textbf{GPU-Specific Code Optimizations}: Predicated execution, register-bound context mapping and expression simplification to reduce warp divergence and improve occupancy.
    \item \textbf{Hierarchical Parallelization}: Two complementary GPU optimizations—\textit{sibling-splitting} and \textit{multi-offload} that expose parallelism across the MG-tree hierarchy.
\end{enumerate}

\section{Background}\label{sec:bg}
\subsection{Problem definition}
\label{subsec:problem_def}

\begin{figure}
\centerline{\includegraphics[width=\linewidth]{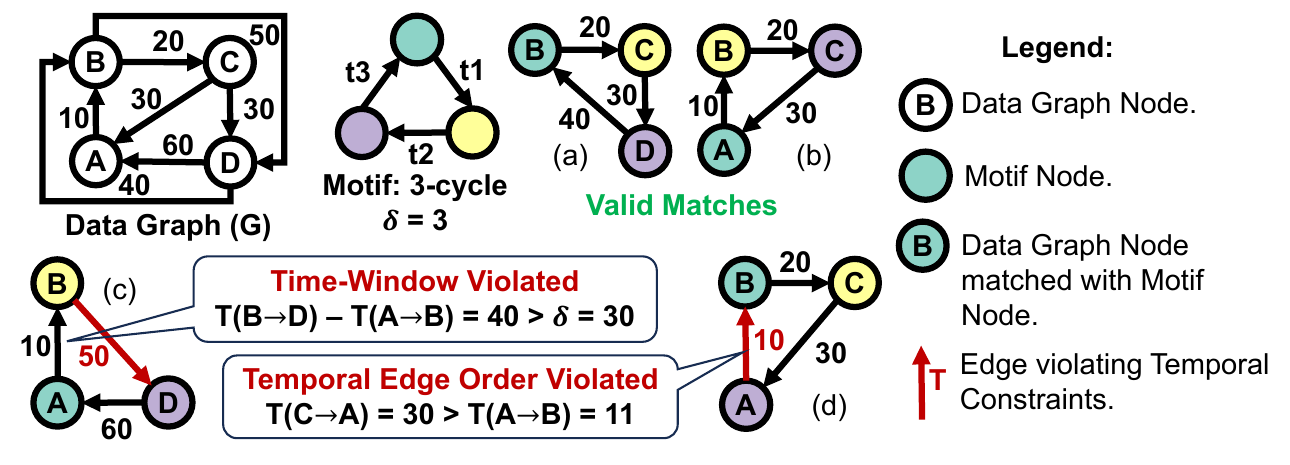}} 
\caption{Example of mining a 3-cycle motif within a temporal Data Graph (G). Data vertices (e.g.~\img{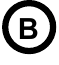}) in the matches are color-coded (\img{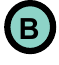}) to their corresponding motif vertex(\img{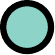}).}
\label{fig:walkthrough_example}
\end{figure}

\textbf{A Temporal Graph} is defined as an ordered collection of \textit{temporal edges}, where each \textit{temporal edge} is a directed connection between two vertices with an associated timestamp.
Formally, a temporal graph $G=(V_G,E_G)$ is a set of vertices $V_G$ connected by a list of $n$ temporal edges $E_G = {(u_i, v_i, t_i)}^n_{i=1} | u_i,v_i \in V_G$, where $u_i$ and $v_i$ are the source and destination vertices of the edge $(u_i, v_i)$, respectively, and $t_i \in \mathbb{R}^+$ is the edge's timestamp.
The edges are chronologically ordered with unique timestamps.
Additionally, both vertices and edges have optional discrete or continuous attributes (\textit{e.g.,} node/edge types).

\noindent
\textbf{A $\delta$-Temporal Motif}
$M=(V_M,E_M)$ is defined as an ordered sequence of $m$ edges, where $E_M = {(u_i, v_i, t_i)}^m_{i=1} | u_i,v_i \in V_M$, that occur within a specified time window of length $\delta \in \mathrm{R}^+$~\cite{mackey2018chronological}.
The edges of $G$ that match with $M$ must be temporally ordered $(t_1 < t_2 ... < t_m)$ and the entire sequence must occur within a time-window $\delta$, i.e.\  $(t_m - t_1 \leq \delta)$.
The label of a motif-edge in all figures indicates the edge's relative order.

\textbf{Temporal Motif Mining}
is the task of mining instances of a $\delta$-temporal motif within a temporal graph.
This process can yield two types of results: either a comprehensive list of all matching motifs (enumeration) or a tally of their occurrences (counting).
Our formulation of this problem requires finding one-to-one correspondences between the vertices of motif and subgraphs within the temporal graph being mined, which is also known as isomorphism-based mining~\cite{everest}.
While some related work in the field of temporal mining adopt slightly different notions of a match, \textit{e.g.,} mapping sets of edges of the temporal graph to the edges of the motif~\cite{Kosyfaki2018FlowMI}, sequences of time-stamped events across snapshots of static networks ~\cite{tm-miner} or homomorphism between edges of the motif and a match ~\cite{Cai2023MobilityAwareOS}, we restrict the scope of this work to a strict isomorphism-based structural definition of a match based on prior works~\cite{everest,mackey2018chronological,paranjape2017motifs}.
We therefore use ``mining'' to denote identifying instances of a pattern, as opposed to very early literature that equated ``mining'' with discovering novel patterns~\cite{chen1996datamining}.

The problem of mining $\delta$ temporal motifs is similar to SQL query execution in traditional databases, where the temporal graph can be thought of as a database and the motif and time-window are analogous to the query.
Temporal mining can then be expressed as nested \texttt{JOIN} operations to match edges in the data graph to edges in the motif by using a vertex as the common key, and \texttt{WHERE} clauses to filter out edges that violate temporal constraints.

\noindent
\textbf{The Walk-Through Example} in Fig.~\ref{fig:walkthrough_example} illustrates the different aspects of temporal motif mining introduced above.
The temporal data graph $G$ has four vertices, $A$ through $D$, and seven edges marked with timestamps using the same unit, e.g., seconds.
The motif's vertices are color-coded to map with their corresponding data graph vertex in a match, with the timestamp of edges indicating their relative temporal order: $T$(\img{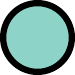}$\rightarrow$\img{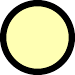}) < $T$(\img{figures/node_b.pdf}$\rightarrow$\img{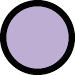}) < $T$(\img{figures/node_c.pdf}$\rightarrow$\img{figures/node_a.pdf}).
Given the temporal graph $G$, a 3-cycle motif to be mined and a time-window $\delta = 30$,  Fig.~\ref{fig:walkthrough_example}(a)-(d) illustrates different sub-graphs that are valid and invalid matches, along with the rationale for their classification.
The difference in the timestamp between the first and the last edges is $2$ for \ref{fig:walkthrough_example}(a) and \ref{fig:walkthrough_example}(b), making them valid matches ($2<\delta=30$).
Fig.~\ref{fig:walkthrough_example}(c) is an invalid match because its second edge occurs more than 30 time-steps after the first edge, placing it outside the permissible time-window.
\ref{fig:walkthrough_example}(c) violates the time-window since the second and third edges occur more than 30 time-steps after the first edge.
Fig.~\ref{fig:walkthrough_example}(d) violates the temporal ordering constraint: $T$(\img{figures/node_b.pdf}$\rightarrow$\img{figures/node_c.pdf}) $=30$ $>$ $T$(\img{figures/node_c.pdf}$\rightarrow$\img{figures/node_a.pdf}) $=10$.

\subsection{Algorithmic Prior Work} \label{subsec:prior-work}

\begin{algorithm}[]
    \scriptsize
    \caption{Temporal Motif Mining Algorithm}
    \label{algo:tm}
    \begin{algorithmic}[1]
        \State $\textbf{Inputs}:\text{Temporal Data Graph G},\text{Motif M}:\text{and} \text{Time-Window size}\:\delta$
\State $\textbf{Output Variables}:\text{count} \gets 0$;
$\text{matches} \gets \emptyset$;
\State $\textbf{Book-Keeping Context}:$\label{algo:tm:bkvar}
$\text{e\_stack}[:] \gets \emptyset$;
$\text{m2g}[:] \gets -1$;
$\text{g2m}[:] \gets -1$;
$\text{incnt}[:] \gets 0$; 

\Statex

\Procedure{\FuncDef{TemporalMining}}{$TemporalGraph\:\text{G},Motif\:\text{M},Int\:\delta$}
\State \Call{\FuncCall{MatchEdge}}{$\text{G}, \text{M}, \delta, 0$}$;$ \label{algo:tm:first_call}
    \Return ${\text{count}, \text{matches}};$ \Comm{Start by matching first motif edge.} 
\EndProcedure

\Statex

\Function{\FuncDef{MatchEdge}}{$TemporalGraph\:\text{G},Motif\:\text{M},Int\:\delta, Int\:e_M$}
    \If{$e_M = |E(\text{M})|$} \Comm{Match Found when $e_M$ reaches number of motif edges.}
        \State $\text{count} \gets \text{count} + 1;$
            $\text{matches}.append(\text{e\_stack});$
            \Return
            \label{algo:tm:match_found}
    \EndIf
    \State $u_M, v_M \gets M.\text{edges}[e_M];$
        $u_G \gets \text{m2g}[u_M];$
        $v_G \gets \text{m2g}[v_M]$
        \label{algo:tm:get_mapping}
        \Comm{Get $u_G\leftrightarrow u_M$ map.}
    \State $\text{cands} \gets (u_G \neq -1)\,?\,\mathcal{N}(u_G):\text{G}.edges$
        \label{algo:tm:cand_selection}
        \Comm{All edges are candidates for unmapped $u_G$.}
    \For{$\text{edge}_G \in \text{cands} $} \label{algo:tm:cand_search}
        \If{$e_M>0\:\textbf{and}\:(\text{edge}_G.t < \text{e\_stack}[e_M\!-\!1].t$ \textbf{or} $\text{edge}_G.t - \text{e\_stack}[0].t > \delta)$}
            \State \textbf{continue}
            \label{algo:tm:temp_const}
            \Comm{Ensure Temporal Edge-Order and Time-Window Constraints.}
        \EndIf
        \If{$v_G \neq -1$ \textbf{and} $\text{edge}_G.v \neq v_G$}
            \textbf{continue}
            \label{algo:tm:struct_const}
            \Comm{Ensure Structural Constraints.}
        \EndIf
        \State $\text{e\_stack}.append(\text{edge}_G);$
        \State \Call{\FuncCall{RollOnEdge}}{$u_M, v_M, \text{edge}_G.u, \text{edge}_G.v$}
            \label{algo:tm:roll_up_edge}
            \Comm{Book-keep $(u_M,v_M)$ to edge$_G$ map.}
        \State \Call{\FuncCall{MatchEdge}}{$\text{G}, \text{M}, \delta, e_M + 1$} 
            \label{algo:tm:recursive_call}
            \Comm{Expand Search-Tree to match next M edge recursively.}
        \State $\text{e\_stack}.pop();$
            \Call{\FuncCall{RollBackEdge}}{$\text{edge}_G.u,\text{edge}_G.u$}$;$
            \label{algo:tm:roll_back_edge}
            \Comm{Remove edge$_G$'s mapping.}
    \EndFor
\EndFunction

\Statex

\Function{\FuncDef{RollOnEdge}}{$Int\:u_M,Int\:v_M,Int\:u_G,Int\:v_G$}
    \State $\text{m2g}[u_M] \gets u_G;$
        $\text{g2m}[u_G] \gets u_M;$
        $\text{m2g}[v_M] \gets v_G;$
        $\text{g2m}[v_G] \gets v_M$
    \State $\text{incnt}[u_G] \gets \text{incnt}[u_G] + 1;$
        $\text{incnt}[v_G] \gets \text{incnt}[v_G] + 1$
\EndFunction

\Statex

\Function{\FuncDef{RollBackEdge}}{$Int\:u_G,Int\:v_G$}
    \State $u_G, v_G \gets \text{edge};$
        $\text{incnt}[u_G] \gets \text{incnt}[u_G] - 1$
        $\text{incnt}[v_G] \gets \text{incnt}[v_G] - 1$
    \If{$\text{incnt}[u_G] = 0$}
        $u_M \gets \text{g2m}[u_G];$
            $\text{g2m}[u_G] = \text{m2g}[u_M] = -1$
    \EndIf
    \If{$\text{incnt}[v_G] = 0$}
        $v_M \gets \text{g2m}[v_G];$
            $\text{g2m}[v_G] = \text{m2g}[v_M] = -1$
    \EndIf
\EndFunction  
    \end{algorithmic}
\end{algorithm}

\begin{figure}[]
  \centering
  \includegraphics[width=\linewidth]{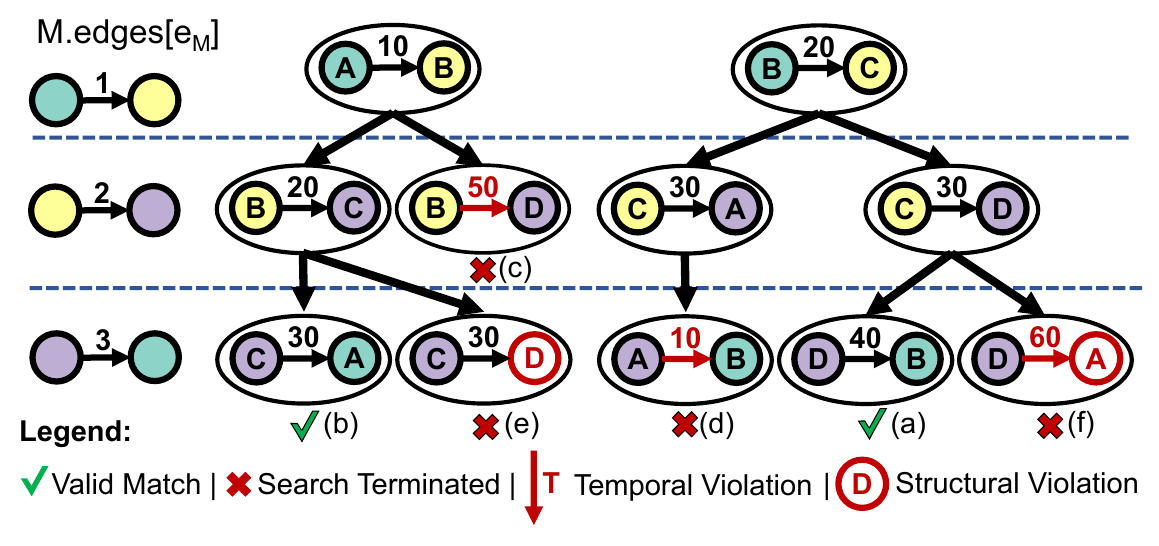}
  \vspace{-8mm}
  \caption{Search-Tree to mine a 3-cycle in Data Graph G in Fig.~\ref{fig:walkthrough_example}. Labeled leaf nodes (a-d) are search-paths of corresponding valid/invalid matches in Fig.~\ref{fig:walkthrough_example}.}
  \label{fig:exmpl_search_tree}
\end{figure}

Temporal motif mining approaches fall into two categories: exact algorithms that enumerate all matches through subgraph isomorphism with temporal constraints~\cite{paranjape2017motifs,everest,mackey2018chronological}, and approximate methods that estimate counts via sampling or sketching~\cite{teacups,edgesampling,sarpe2021presto,liu2019sampling,sarpe2021oden}.
While approximate techniques scale better for large graphs, exact methods remain crucial for applications requiring complete enumeration, such as fraudulent activity identification in financial networks~\cite{hajdu2020temporal} or insider threat identification~\cite{glasser2013bridging,mackey2018chronological}.
Notably, many approximate methods still leverage exact algorithms as subroutines for local pattern matching~\cite{liu2018sampling,sarpe2021presto}.
Given these considerations, \textit{our work  focuses on exact algorithms}, with a brief discussion of approximate techniques deferred to \S~\ref{sec:related-work}.


Paranjape \textit{et al.}~\cite{paranjape2017motifs} formalized $\delta$-temporal motif mining and proposed a two-phase algorithm: 1) enumerating static isomorphic subgraphs, then 2) verifying temporal constraints.
The first step uses existing static subgraph / pattern mining methods that identify matches that are structurally equivalent to the motif, ignoring temporal constraints.
This potentially leads to unnecessary computational overhead, as structurally compliant candidates may not satisfy temporal requirements~\cite{everest}.
Subsequent work by Mackey \textit{et al.}~\cite{mackey2018chronological} improved efficiency by pruning temporally invalid candidates before expanding the entire subgraph.
Everest~\cite{everest} adapted this approach for GPUs with a state-of-art warp-level parallelization of candidate exploration.
The essence of Mackey's algorithm~\cite{mackey2018chronological} is captured in Algorithm~\ref{algo:tm}, with a subsequent discussion on Everest's~\cite{everest} distinctions from this approach.

\noindent
\textbf{Data-Structures:} The algorithm employs a set of book-keeping variables (line~\ref{algo:tm:bkvar}) to map a motif edge and a graph edge.
Specifically, "m2g" and "g2m" facilitate bidirectional mapping between motif and graph vertices, "e\_stack" maintains a temporally ordered stack of all currently matched edges, while "incnt" tracks the number of matched edges incident on each graph vertex.

\noindent
\textbf{Search-Tree:} Mining the graph can be conceptualized as a tree, where nodes within a level represent candidate edges for the corresponding motif edge, and the parent corresponds to the graph edge matched to the preceding motif edge.
Fig.~\ref{fig:exmpl_search_tree} illustrates a few search-trees for mining a 3-cycle in the data graph G (Fig.~\ref{fig:walkthrough_example}), with individual paths labeled to correspond to matches Fig.~\ref{fig:walkthrough_example}(a-d).

Algorithm~\ref{algo:tm} begins by mapping the first motif edge (line~\ref{algo:tm:first_call},$e_M=0$), considering all edges in the data graph as potential candidates due to the absence of pre-existing vertex mappings.
It then employs a depth-first exploration strategy to recursively expand the search tree for subsequent motif edges (line~\ref{algo:tm:recursive_call},$e_M>0$).
When exploring a new level, the algorithm prunes the candidate list to the out-edges of a potentially mapped source vertex $u_G$.
The candidates are further pruned based on temporal and structural constraints (line~\ref{algo:tm:temp_const}-\ref{algo:tm:struct_const}).
Fig.~\ref{fig:exmpl_search_tree}(c,d) illustrate the search tree being pruned due to temporal violations, corresponding to the invalid matches Fig.~\ref{fig:walkthrough_example}(c,d).
The structural constraint enforces a bijective mapping between the destination vertex, $v_M$, and graph vertex, $v_G$.
Search paths terminating at Fig.~\ref{fig:exmpl_search_tree}(e,f) illustrate structural violations since the last edge must finish the cycle by ending at the first vertex ($A$ for (e) and $B$ for (f)).
After passing the constraint checks, the book-keeping context is updated to capture the mapping between $e_M$ and $\text{edge}_G$ (line~\ref{algo:tm:roll_up_edge}), and recursively proceeds to the next motif edge.
Once all motif edges have been matched (Fig.~\ref{fig:exmpl_search_tree}(a,b)), the algorithm records the match by either incrementing a counter or adding it to an enumeration list (line~\ref{algo:tm:match_found}).
This algorithm can be parallelized over candidates for the first edge, but becomes challenging for subsequent edges due to the sequential nature of candidate generation (line~\ref{algo:tm:cand_search}).
Everest~\cite{everest} extends Algo.~\ref{algo:tm} to GPUs by storing the candidate list as a range of edges.
By distributing these ranges among threads within a GPU warp, Everest~\cite{everest} enables a massively parallel exploration of candidates across all levels of the search-tree.

\section{A Case for Multi-Query Execution}\label{sec:motiv}


This section motivates the need for efficient multi-query execution and evaluates the effectiveness of existing techniques in addressing the challenges of temporal motif mining.


Modern analytical workloads increasingly require concurrent execution of multiple queries across domains ranging from financial fraud detection to social network analysis.
Anti-money laundering systems must simultaneously track diverse transaction patterns \cite{blanuvsa2024graph,altman2023realistic,Sommers2024,Fisher2024}, while social platforms analyze user engagement through parallel interaction queries \cite{BalanRege2017,ZACHLOD20221064}.
Traditional single-query optimization proves inadequate for these workloads due to \textit{redundant computations across overlapping queries}.

\subsection{Collective Queries in Traditional Databases}

The challenge of multi-query optimization (MQO) has been studied since Sellis' seminal work on identifying common subexpressions \cite{sellis1988multiple}, though optimal planning remains NP-hard \cite{sellis1990multiple}.
Contemporary approaches fall into two categories: 1) physical optimizations that make the underlying system more efficient (\textit{e.g.,} shared data access patterns via scan-sharing \cite{candea2009scalable,arumugam2010datapath} or resource-aware scheduling \cite{boncz2006monetdb,agarwal2013blinkdb}), and 2) algebraic transformations that algorithmically reduce the amount of work~\cite{tu2022multi}.
Ren and Wang~\cite{ren2016multi} pioneered MQO for isomorphic-pattern mining, by reducing redundant computations by identifying shared structures across queries.
Subgraph Morphing~\cite{jamshidi2023accelerating}, a hybrid approach combining algebraic and physical optimizations, decomposes query patterns into alternative patterns that are less computationally expensive to mine.

\subsection{Opportunity: Commonality in Temporal Motif Queries}\label{subsec:opportunity}

Building upon insights gleaned from prior work, we explore potential ways to identify and reduce redundant computation.
The techniques proposed previously ~\cite{ren2016multi,jamshidi2023accelerating} could be adapted to exploit structural similarities among temporal queries, but not without challenges.
Ren and Wang~\cite{ren2016multi} can be used to first identify isomorphic-matches with query motifs and then filtered  to comply with temporal constraints, leading to unnecessary exploration of a large search-space ~\cite{paranjape2017motifs}.
Subgraph Morphing~\cite{jamshidi2023accelerating} exploits symmetry in graph isomorphism to reduce redundant searches, may produce matches that violate temporal constraints in the context of temporal motif mining (\textit{e.g.,} Fig.~\ref{fig:walkthrough_example}(b) and (d)).
This issue is exacerbated when two vertices have multiple edges between them, leading to a combinatorial explosion in the number of potential matches requiring a large memory footprint for enumeration.

\begin{figure}[]
  \centering
  \includegraphics[width=\linewidth]{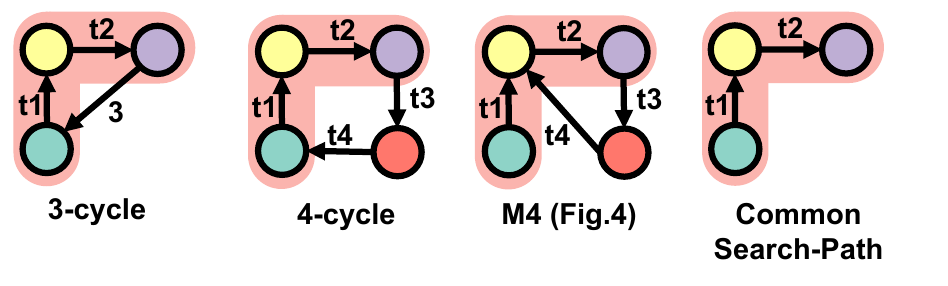}
  \vspace{-8mm}
  \caption{Common search-path between 3-cycle, 4-cycle and M4 motifs.}
  \label{fig:common_path}
\end{figure}

Given the challenges associated with efficiently adapting existing techniques for temporal motif mining, we allude to an approach specifically tailored to this domain.
Consider Fig.~\ref{fig:common_path}, which illustrates the 3-cycle, 4-cycle, and ~\textt{M4} motifs (Fig.~\ref{fig:walkthrough_example}).
Observe that the motifs share the same first two edges, \img{figures/node_a.pdf}$\xrightarrow{\text{\scriptsize{t1}}}$\img{figures/node_b.pdf} and \img{figures/node_b.pdf}$\xrightarrow{\text{\scriptsize{t2}}}$\img{figures/node_c.pdf}, indicating that the ideal mining algorithm must visit these edges in the search-tree for either a 3-cycle or a 4-cycle before proceeding to expand to the third edge.
This observation suggests that we can eliminate redundant computation along the common path.
By prioritizing edge traversal in a chronological order, we establish a natural heuristic that favors exploring common paths before diverging into specific motif searches.
In subsequent section, we generalize this concept to accommodate more temporally and structurally complex common paths, laying the foundation for our efficient co-mining approach.

\section{\name~Design}\label{sec:design}

This section presents the design of \name, outlining its core objectives, workflow, key algorithmic components and the runtime for CPU/GPU backend.
We introduce the Motif-Group Tree (MG-Tree) generation algorithm and the co-mining algorithm, which form the foundation of our approach.

\subsection{Design Goals}

\name~is designed with three primary objectives to address the challenges in multi-query execution for temporal motif mining.

\textbf{Efficient Motif Co-Mining.}
While existing approaches to temporal motif mining have focused on optimizing mining a single motif, our observation in \S\ref{subsec:opportunity} presents an opportunity to identify redundant computations when mining multiple motifs. 
We aim to minimize these redundant computations by exploiting structural and temporal similarities among motifs within the query set, reducing the overall workload associated with mining multiple motifs.

\textbf{Multi-Backend Support.}
To ensure broad accessibility, one of our goals is to support seamless operation on both GPU and CPU platforms.
Recognizing the memory constraints associated with GPUs and that not all users have access to high-performance GPUs, our design allows users the flexibility to execute motif mining tasks on CPU resources when necessary.

\textbf{High Performance Optimizations.}
While co-mining presents clear opportunities for performance improvements, its implementation introduces specific challenges that must be addressed.
For instance, the additional context and control-flow introduced to enable co-mining on GPU threads can potentially reduce performance due to the additional register footprint and warp-divergence from threads in the same warp mining different motifs.
Our goal is to address these challenges (\S\ref{sec:opt}), ensuring that the system not only capitalizes on the benefits of co-mining but also maintains high performance throughout the mining process.

\subsection{\name~Workflow}

\begin{figure}
  \centering
  \includegraphics[width=0.7\linewidth]{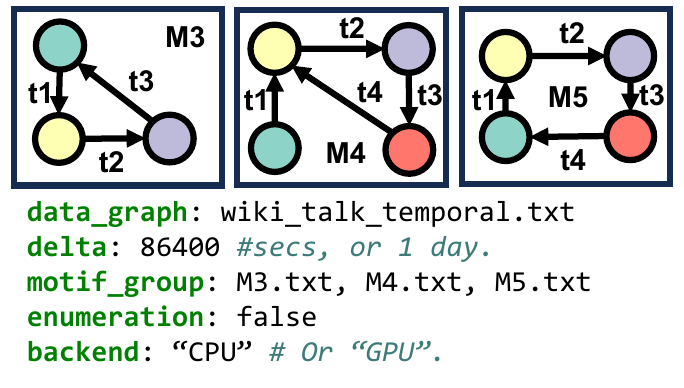}
  \vspace{-0.25cm}
  \caption{Example User Query.}
  \label{fig:exmpl_query}
\end{figure}

\begin{figure}
  \centering
  \includegraphics[width=0.8\linewidth]{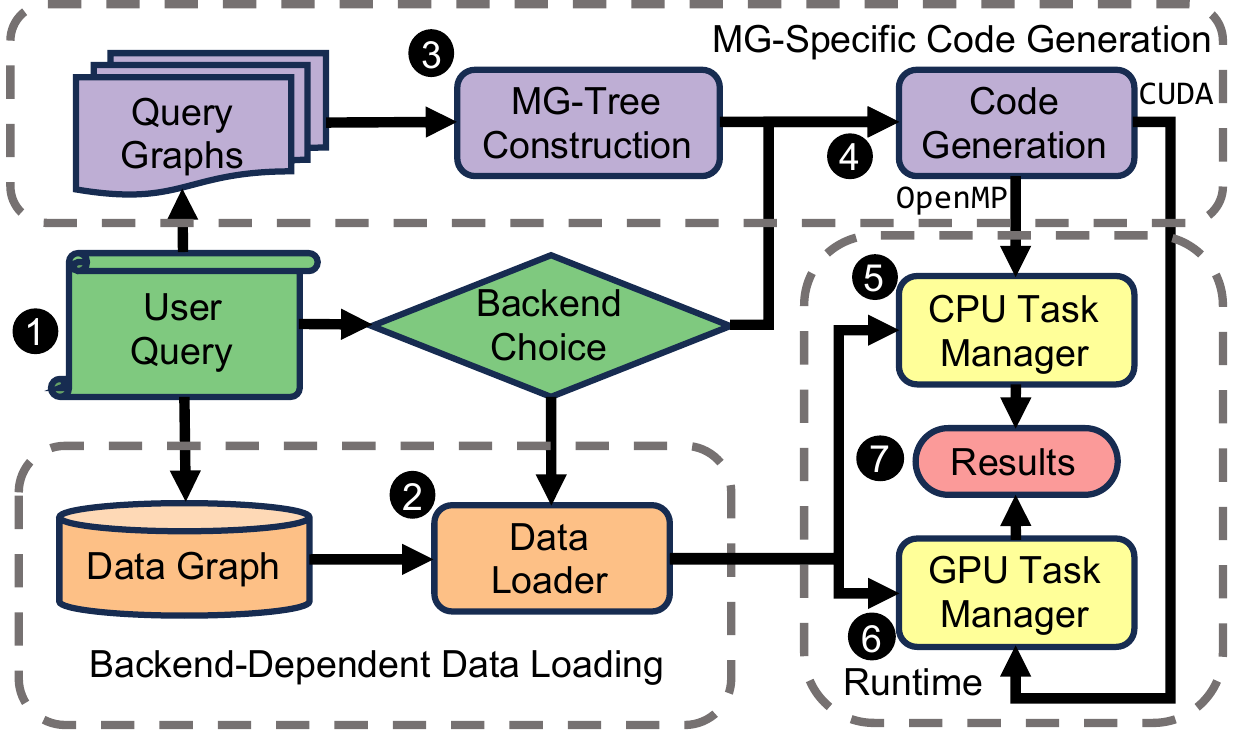}
  \caption{Overview of \name's workflow.}
  \label{fig:workflow}
\end{figure}

Our system takes as input a user query (Fig.~\ref{fig:exmpl_query}) that specifies the temporal data graph to mine (Fig.~\ref{fig:workflow}~\circled{1}), a group of motifs (\textit{i.e.,} motif group), time-window $\delta$, backend choice (CPU or GPU), and whether matches need to be counted or enumerated.
The output is either the per-motif count or enumeration.
The workflow of~\name~is structured into three distinct phases: Compile-Time, Data-Loading and Runtime, visualized by Fig.\ref{fig:workflow}.
The Compile-Time phase generates and compiles the code that implements the co-mining algorithm, while the Data-Loading phase operates concurrently to load the dataset into (CPU/GPU) memory.
The Runtime phase executes the compiled code to mine the loaded dataset.

\noindent
\textbf{Data-Loading and Compile-Time Phases} consists of mechanisms for 1) Data Preprocessing \circled{2}, 2) Motif-Group Tree (MG-Tree) Construction  \circled{3}, and 3) Backend-Specific Code Generation and Compilation~\circled{4}.
The temporal data graph specified by the query \circled{1} is preprocessed from an edge-list format ($u_i,v_i,t_i$) to an adjacency-list format stored in a CSR-like structure, with edges sorted in ascending order of timestamps~\circled{2}.
The system constructs a hierarchical representation of the query motifs in the motif group, \textit{i.e.,} the MG-Tree, capturing structural and temporal similarities~\circled{3}.
The system then generates optimized code tailored to the specific MG-Tree and backend, utilizing C++ with OpenMP for CPU execution or CUDA C for GPU execution~\circled{4}.

\noindent
\textbf{Runtime Phase} dispatches the compiled code and preprocessed data to the appropriate backend's task manager (\circled{5}
 for CPU,~\circled{6} for GPU).
To address the inherent load imbalance in graph workloads, both CPU and GPU task managers implement sophisticated load balancing strategies~(\S\ref{subsec:runtime}).
Upon completion, the system aggregates the results, providing either motif counts or enumerations as specified in the query~\circled{7}.

\subsection{Motif-Group Tree Construction}

We propose a hierarchical data structure called Motif-Group Tree (MG-Tree) that captures structural and temporal similarities among query motifs to facilitate efficient co-mining.
By grouping motifs that share overlapping edges (with the same relative order) in a hierarchical structure, the MG-Tree enables the co-mining algorithm to reuse computations along the shared search paths and eliminating redundant work.
This hierarchical structure  also exposes algorithmic parallelism to improve concurrency.
While the MG-Tree is not the first to exploit structural similarities to accelerate matching~\cite{wang2020ferrari,jin2012prague}, we are the first to propose such a solution for temporal mining.
We define the MG-Tree by defining its composition and the constraints that define the relationship between parent and children nodes,

\noindent
\textbf{\textsc{MG-Tree Definition}}:
For a group of temporal motifs $MG = \{M_1, M_2, \ldots, M_k\}$, the MG-Tree $MGT$ is defined as a tree of Nodes that capture the similarities among motifs in $MG$, rooted at $N_{root}$.

\textbf{Node Composition:} Any Node $N \in MGT$ is composed of the following 3 members: $C_N$, Children($N$) and $Q_N$,

\begin{description}[leftmargin=5pt]
    \item[$C_N$\textbf{:}] A motif with edges common across all descendants $C_{N_\text{desc}}$, \textit{i.e.} a prefix for $C_{N_\text{desc}}$ with their first $|C_N|$ edges being equal to $C_N$.
        
    \item[\( N_\text{child} \in \mathrm{Children}(N) \)\textbf{:}] Immediate descendants constructed by extending $C_N$, where $C_N$ is the longest non-trivial prefix of $C_{N_\text{child}}$.
        
    \item[$Q_N$\textbf{:}] The reference to a query motif $M_i \in MG$ when $C_N$ is equivalent to $M_i$, else is $\emptyset$.
    $\forall M_i \in MG, \exists N \in MGT~|~Q_N = M_i$

\end{description}

\textbf{Root Node:} $N_{root} \in MGT$ whose common motif $C_{N_{\text{root}}}$ has edges common across all motifs in $MG$.


\textcolor{blue}{
}

The MG-Tree construction algorithm (Algo.~\ref{algo:mg_tree_const}) begins by invoking the \textsc{ConstructMGTree} procedure on the motif group $MG$.
For the sake of brevity and simplicity, we refer to the temporal order of a motif edge as its timestamp.
A TMap is generated for each motif, mapping timestamps to the corresponding edges.
Nodes with children are known as Intermediate Nodes (Intr.), and those without are Leaf Nodes.
Members of the motif group are populated as leaf Nodes in the MG-Tree, with their $C_N$ and $Q_N$ references set to the member (line \ref{algo:mg_tree_const:insert_motif}).
Upon reaching a leaf Node N during co-mining, the search is limited to mine only $Q_N$.
The algorithm then proceeds in a recursive manner to build the MG-Tree, starting from the first edge in all motifs (line \ref{algo:mg_tree_const:first_edge}).
The motifs are grouped together based on the source and destination of the edges at timestamp $T$ (line \ref{algo:mg_tree_const:edge_grouping}).
Motifs in singleton groups are inserted as a list of children into their parent Node ($p\_gid$).
Undivided groups, \textit{i.e.,} all motifs in the input group have the same edge at $T-1$ and $T$, end up reusing the node created at $T-1$ (or potentially before) (line \ref{algo:ctm:reusing_gid}).
A new Intr. Node is created for all other groups, representing motifs encountered during the search process but not necessarily counted or enumerated like $Q_N$.
The algorithm also eliminates redundant work when $C_N$ is identical to a $Q_N$ in the child\_group.
The new Intr. Node is then added as a child to its parent (line \ref{algo:ctm:insert_child}).
The MG-Tree for each child group is recursively constructed and attached to the final MG-Tree~(line \ref{algo:mg_tree_const:recursive_call}), with the recursion terminating at Leaf Nodes.

\begin{figure}
    \centering
    \begin{minipage}{0.53\linewidth}
        \centering
        \includegraphics[width=\textwidth]{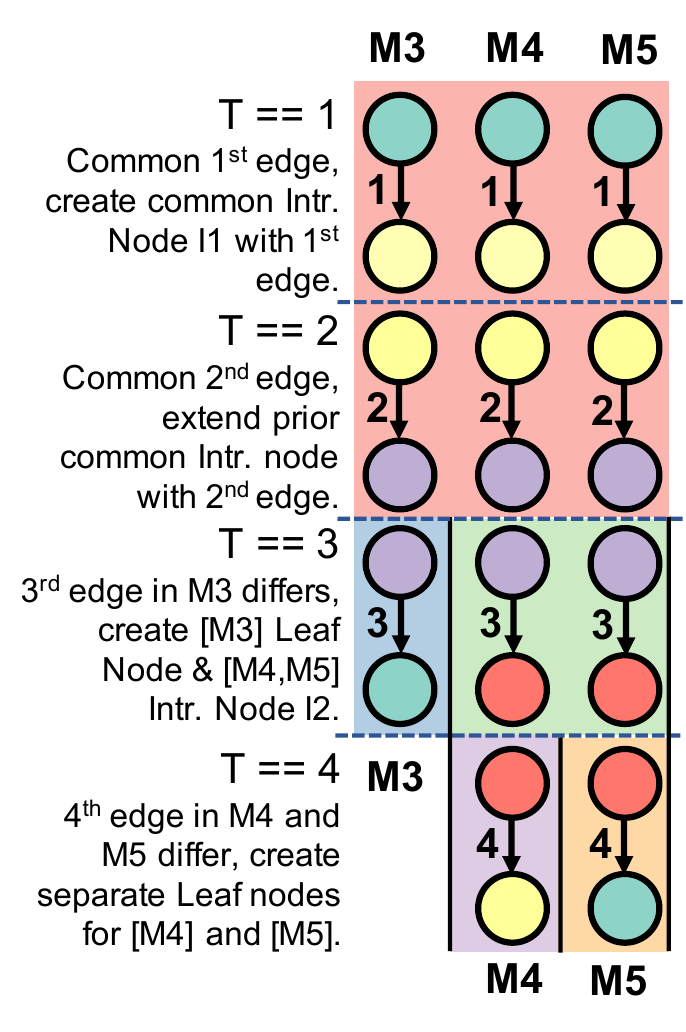}
        \caption{Visualizing MG-Tree Construction.}
        \label{fig:viz_mg_tree_cosnt}
    \end{minipage}\hfill%
    \begin{minipage}{0.45\linewidth}
        \centering
        \includegraphics[width=\textwidth]{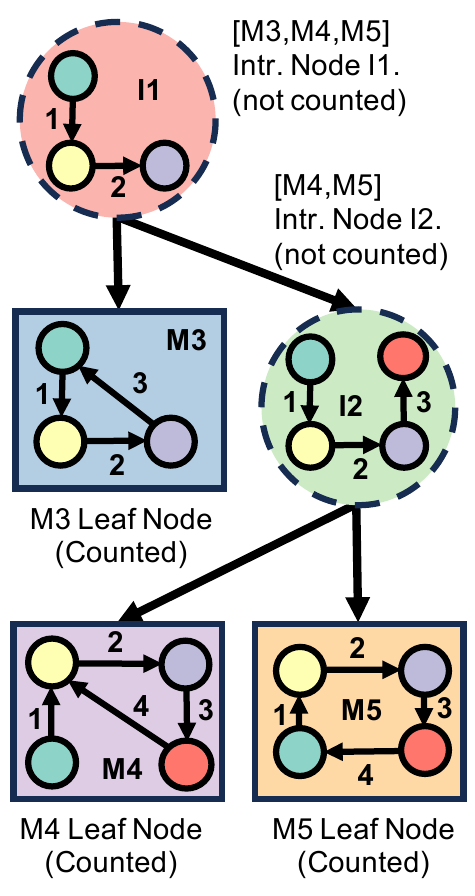} 
        \caption{MG-Tree of motifs in Fig.~\ref{fig:exmpl_query}.}%
        \label{fig:mg_tree_exmpl_query}
    \end{minipage}
\end{figure}

\textbf{Walkthrough Example:}
Consider the motif group [\textt{M3},\textt{M4},\textt{M5}] from Fig.~\ref{fig:exmpl_query} as an input to the algorithm.
Fig.~\ref{fig:viz_mg_tree_cosnt} visualizes the construction of the MG-Tree in Fig.~\ref{fig:mg_tree_exmpl_query}. 
After setting up TMap and the MG-Tree with leaf Nodes, \textsc{CreateTree} is invoked on the motif group.
Since all motifs have the same edge at $\text{T} = 1$, \textit{i.e.,} \img{figures/node_a.pdf}$\rightarrow$\img{figures/node_b.pdf}, the child\_group reuses $N_\text{root}$ at $\text{c\_gid} = 0$, referred to as Intermediate Node \textt{I1}.
$N_\text{root}$'s $C_N$ is set to contain the single edge, \img{figures/node_a.pdf}$\rightarrow$\img{figures/node_b.pdf}, and the $Q_N$ is left empty since none of motifs resemble $C_N$.
With the recursive call to \textsc{CreateTree}, all the motifs end up being grouped together at $\text{T} = 2$ as well since they still share an edge, \img{figures/node_b.pdf}$\rightarrow$\img{figures/node_c.pdf}.
The $C_N$ for \textt{I1} is reset to \img{figures/node_a.pdf}$\rightarrow$\img{figures/node_b.pdf}$\rightarrow$\img{figures/node_c.pdf}.
At $\text{T} = 3$, \textt{M3} is grouped separately from \textt{M4} and \textt{M5} since its edge, \img{figures/node_c.pdf}$\rightarrow$\img{figures/node_a.pdf}, is different from that of \textt{M4} and \textt{M5},\img{figures/node_c.pdf}$\rightarrow$\img{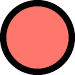}.
\textt{M3} is added as a child for \textt{I1}.
A new Intr. Node, \textt{I2},  is created for \textt{M4} and \textt{M5} with the $C_N$, \img{figures/node_a.pdf}$\rightarrow$\img{figures/node_b.pdf}$\rightarrow$\img{figures/node_c.pdf}$\rightarrow$\img{figures/node_d.pdf}.
With \textt{I2} as a parent Node, \textsc{CreateTree} is called on [\textt{M4},\textt{M5}] and ends up separating them since they have different edges at $\text{T} = 4$, and are added as children to \textt{I2}.
Observe that if we had to build an MG-tree only for [\textt{M4},\textt{M5}], it would have been the tree rooted at \textt{I2}.
This hierarchical representation enables the exploitation of similarities between motifs at various granularities, reducing the overall computational workload.

\begin{algorithm}[]
    \scriptsize
    \caption{MG-Tree Construction}
    \begin{algorithmic}[1]
        \Statex \textbf{Initialize Context}
\State $\text{TMap} \gets \emptyset$; $\text{mg_tree} \gets \emptyset$; $\text{unique\_gid} \gets 0$

\Statex \textbf{Input:} List of unique motifs; \textbf{Output:} MG-Tree $MGT$

\Procedure{\FuncDef{ConstructMGTree}}{$List[Motif]$ MG}
    \For{$M \in MG$} \Comm{Generate tmaps and insert tree-nodes}
        \State $\text{TMap}[M] \gets \Call{\FuncCall{GraphToTMap}}{M}$;
        \Call{\FuncCall{InsertMotif}}{M};
    \EndFor
    \State $\text{root\_gid} \gets \Call{\FuncCall{GetNewUniqueGID}()}{}$ \Comm{Create MG\_Tree\_Node at gid 0}
    \State $N_\text{root} \gets MGT[\text{root\_gid}]; \text{Children}(N_\text{root}).\text{clear}(); C_{N_\text{root}} \gets Q_{N_\text{root}} \gets \varnothing$ 
    \State \Call{\FuncCall{CreateTree}}{{1,\Call{\FuncCall{GetNewUniqueGID}()}{},    $MG$}} 
        \label{algo:mg_tree_const:first_edge}
    \State \Return $\text{mg_tree}$
\EndProcedure

\Statex

\Function{\FuncDef{InsertMotif}}{$Motif$ M} \label{algo:mg_tree_const:insert_motif} \Comm{Query-Motifs are MG-Tree's leaves}
    \State $N \gets MGT[M.\text{name}]; C_N \gets Q_N \gets \text{M}; \text{Children}(N) \gets \emptyset$
\EndFunction

\Statex

\Function{\FuncDef{GraphToTMap}}{$Motif$ M} \label{algo:GraphToMap}
    \State $\text{tmap} \gets \emptyset;$ \Comm{Maps specific timestamp to a static edge}
    \For{$e \in E(M)$}
        $\text{tmap}[e.t] \gets (e.u,e.v)$
    \EndFor
    \State \Return $\text{tmap}$
\EndFunction

\Statex

\Function{\FuncDef{GetNewUniqueGID}()}{} \Comm{Generate unique ID number for constituent motif groups}
    \State $\text{new_gid} \gets \text{unique_gid}; \text{unique_gid} \gets \text{unique_gid} + 1; \Return \Call{\FuncCall{Str}}{\text{new_gid}}$
\EndFunction

\Statex

\Function{\FuncDef{CreateTree}}{$Time$ T, $Str$ p\_gid, $List[Motif]$ motif_group}
    \State $\text{edge_group} \gets \emptyset; N_\text{parent} \gets MGT[\text{p\_gid}]$
    \For{$M \in \text{motif_group}$} \Comm{Group graphs based on edge at T}
        \If{$|\text{TMap}[M]| < T$}
            \textbf{pass} \Comm{Ignore graphs that are too small}
        \EndIf
        \State $\text{edges} \gets \text{TMap}[M]; e \gets \text{edges}[T]; \text{edge_group}[e] \gets \text{edge_group}[e] \cup \{M\}$ \label{algo:mg_tree_const:edge_grouping}
    \EndFor

    \For{$(e, \text{child\_group}) \in \text{edge_group}$}
        \State $q_N \gets \varnothing$
        \For{$M \in \text{child\_group}$} \Comm{Check if a constituent query motif}
            \If{$|E(M)| = T$} \Comm{is equivalent to the group's}
                \State $Q_{N_\text{parent}} \gets \text{M}$; \textbf{break} \Comm{parent motif}
            \EndIf
        \EndFor
        
        \If{$|\text{child\_group}| = 1$} \Comm{Singleton groups do not need a}
            \State $\Call{\FuncCall{InsertChild}}{\text{p\_gid}, \text{child\_group[0]}}$ \Comm{new Node.}
        \Else
            \If{$\text{motif_group} = \text{child\_group}$}
                \State $\text{c\_gid} \gets \text{p\_gid}$
                    \label{algo:ctm:reusing_gid}
                    \Comm{Reuse gid if motif_group was not split up}
            \Else
                \State $\text{c\_gid} \gets \Call{\FuncCall{GetNewUniqueGID}()}{}$
            \EndIf
            \State common\_edges $\gets$ TMap[motif\_group[ \Call{\FuncCall{Random}}{1,|\text{motif\_group}|} ]][1:T+1]
            \State $N$ $\gets$ $MGT[$c\_gid$]$ $C_N \gets \Call{\FuncCall{Motif}}{\text{common\_edges}};Q_N \gets \varnothing$ 
            \State $\Call{\FuncCall{InsertChild}}{\text{p\_gid}, \text{c\_gid}};$\Call{\FuncCall{CreateTree}}{T+1,c\_gid,child\_group} \label{algo:mg_tree_const:recursive_call}
        \EndIf
    \EndFor
\EndFunction

\Statex

\Function{\FuncDef{InsertChild}}{$Str$ p\_gid, $Str$ child_id}
    \If{$\text{p\_gid} \neq \text{child\_id}$}
        \label{algo:ctm:insert_child}
        \Comm{Avoid self-loops if motif_group is intact}
        \State $\text{mg\_tree[p\_gid].children} \gets \text{mg\_tree[p\_gid].children}  \cup \{\text{child\_id}\}$
    \EndIf
\EndFunction
    \end{algorithmic}
    \label{algo:mg_tree_const}
\end{algorithm}



\subsection{\name~Co-Mining Algorithm}

The co-mining algorithm extends the temporal motif mining algorithm (Algo.~\ref{algo:tm}) by using the MG-Tree, instead of a single motif, to guide the expansion of the search tree.
The pseudo-code of the algorithm is outlined in Algo~\ref{algo:ctm}, focusing only on its modifications from Algo.~\ref{algo:tm}.
The core mechanism for mining a single motif-edge remains unchanged (lines~\ref{algo:ctm:unmodified_beg}-~\ref{algo:ctm:unmodified_end}).
The algorithm begins by mining $C_{N_\text{root}}$.
Upon detecting a match for $C_N$, it is counted if the Node also represents a query motif (line.~\ref{algo:ctm:chk_query_ref}).
Subsequently, the algorithm moves onto mining the children of $N$ by using $C_N$'s match as a partial match~(line.\ref{algo:ctm:recursive_call}).
By expressing co-mining in this recursive manner, we are able to reuse most of the mechanisms used in Algo~\ref{algo:tm}.

\begin{algorithm}
    \scriptsize
    \caption{Co-Mining Algorithm}
    \begin{algorithmic}[1]
        \Statex $\textbf{Inputs:}\:\text{Data Graph G},\,\text{root of MG-Tree}\:\text{and}\:\text{time-window}\,\delta$
\Statex $\textbf{Outputs:}\:\text{Counts and matches of motifs in MG-Tree.}$

\Statex

\Procedure{\FuncDef{TemporalCoMining}}{$TemporalGraph\:\text{G},Node\:N_\text{root},Int\:\delta$}
\State \Call{\FuncCall{CoMatchEdge}}{$\text{G}, N_\text{root}, \delta, 0$}$;$ \label{algo:ctm:first_call}
    \Return ${\text{count}, \text{matches}};$ \Comm{Start by matching first motif edge.}
\EndProcedure

\Statex

\Function{\FuncDef{CoMatchEdge}}{$TemporalGraph\:\text{G},Node\:N,Int\:\delta,Int\:e_M$}
    \State $M \gets C_N$ \Comm{M is local and does not affect other calls to \textsc{\FuncCall{CoMatchEdge}}}
    \If{$e_M = |E(M)|$} \Comm{Match Found when $e_M$ reaches number of motif edges.}
        \If{$Q_N \neq \varnothing $}
            \label{algo:ctm:chk_query_ref}
            \Comm{Add match when $Q_N$ is assigned}
            \State $\text{count}[Q_N] \gets \text{count}[Q_N] + 1;$
                $\text{matches}[Q_N].\FuncCall{append}(\text{e\_stack});$
                \label{algo:ctm:match_found}
        \EndIf
        \For{${C_{N_\text{child}}} \in \text{Children}(N)$}
            \State $\triangleright$ \textcolor{teal}{Expand search-tree into mining child motif with e\_stack as partial match.}
            \State $\Call{\FuncCall{CoMatchEdge}}{\text{G},{C_{N_\text{child}}},\delta,e_M}$ \Comm{NOTE: Edge $e_M$ is yet to be mined}
        \EndFor
        \State \Return \Comm{After exploring child motifs, find more matches of M to expand the search-tree.}
    \EndIf
    \State $\boldsymbol{\cdots}$ $\triangleright$ \textcolor{teal}{Unmodified} \textsc{\FuncCall{MatchEdge}} \textcolor{teal}{pseudocode from lines ~\ref{algo:tm:get_mapping} to ~\ref{algo:tm:roll_up_edge} in Algorithm~\ref{algo:tm}}
        \label{algo:ctm:unmodified_beg}
    \State \hspace{2.1mm} \Call{\FuncCall{CoMatchEdge}}{$\text{G}, node, \delta, e_M + 1$}
        \label{algo:ctm:recursive_call}
        \Comm{Expand Search-Tree to match next edge in M.}
    \State $\boldsymbol{\cdots}$ $\triangleright$ \textcolor{teal}{Remaining} \textsc{\FuncCall{MatchEdge}} \textcolor{teal}{ and other pseudocode from line ~\ref{algo:tm:roll_back_edge} in Algorithm~\ref{algo:tm}}
        \label{algo:ctm:unmodified_end}
\EndFunction

    \end{algorithmic}
    \label{algo:ctm}
\end{algorithm}

\begin{figure}
    \centering
    \includegraphics[width=0.9\linewidth]{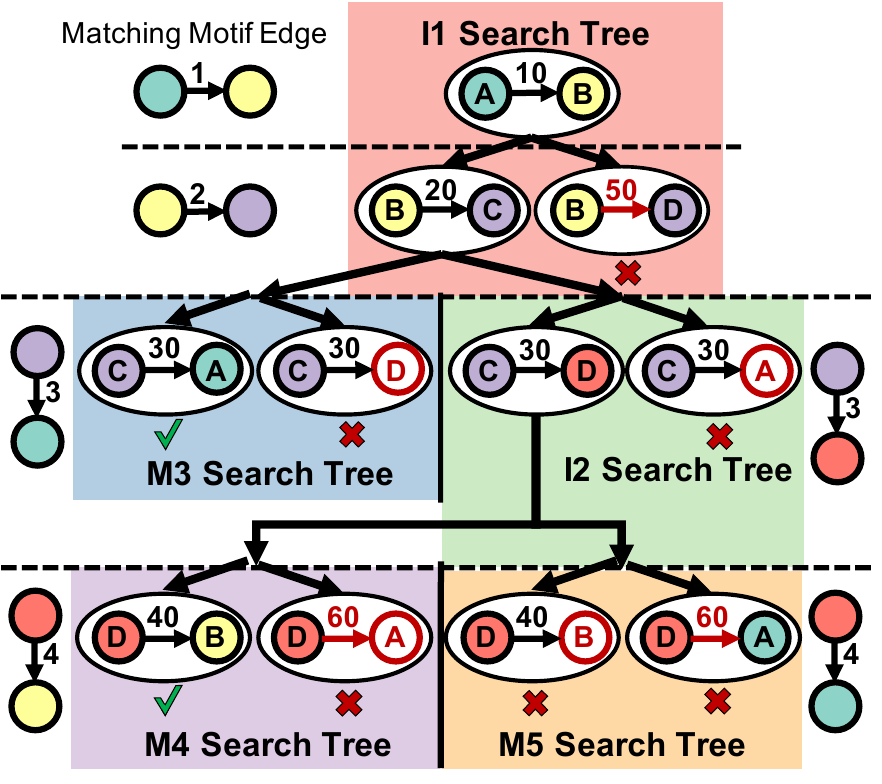} 
    \caption{Search-Tree to mine motifs \textt{M3},\textt{M4} and \textt{M5} reflects the hierarchical structure of their MG-Tree (Fig.~\ref{fig:viz_mg_tree_cosnt}).}
    \label{fig:co_mining_search_tree}
\end{figure}

\noindent
\textbf{Walkthrough Example:} 
Fig.~\ref{fig:co_mining_search_tree} illustrates a small portion of the search-tree explored by the Algo.~\ref{algo:ctm} when mining the motif group [\textt{M3},\textt{M4},\textt{M5}] on the Data Graph G in Fig.~\ref{fig:walkthrough_example}.
The background colors correspond to that of the corresponding motif in the MG-tree (Fig.~\ref{fig:mg_tree_exmpl_query}) being explored by the search-tree.
The motif edges on the sides of the tree at each level indicate the motif edge being mined at that particular level.
Since the search tree is for an MG-Tree, different motif edges can be mined at the same level for different sibling Nodes.
The search tree starts with identifying matches for Node \textt{I1} (top two levels).
With matches of \textt{I1} as partial matches, the search-tree expands into mining the \textt{M3} leaf Node and \textt{I2} Intr. Node, eventually terminating at the leaf Nodes:  \textt{M3},\textt{M4} or \textt{M5}.

\subsection{\name~Runtime} \label{subsec:runtime}

The system initiates the runtime phase by enabling the backend specified by the user.
This backend receives the MG-Tree, $\delta$, and pointers to the temporal data graph located within its memory.
Both CPU and GPU backends parallelize the co-mining algorithm across multiple threads, employing task managers to handle work-scheduling and load-balancing.
While global inputs such as the MG-Tree and $\delta$ are shared, the search-tree context (\textit{e.g.,} vertex maps, counters) is maintained locally within each thread to eliminate contention and maximize parallelism.



\begin{figure}
  \centering
  \includegraphics[width=\linewidth]{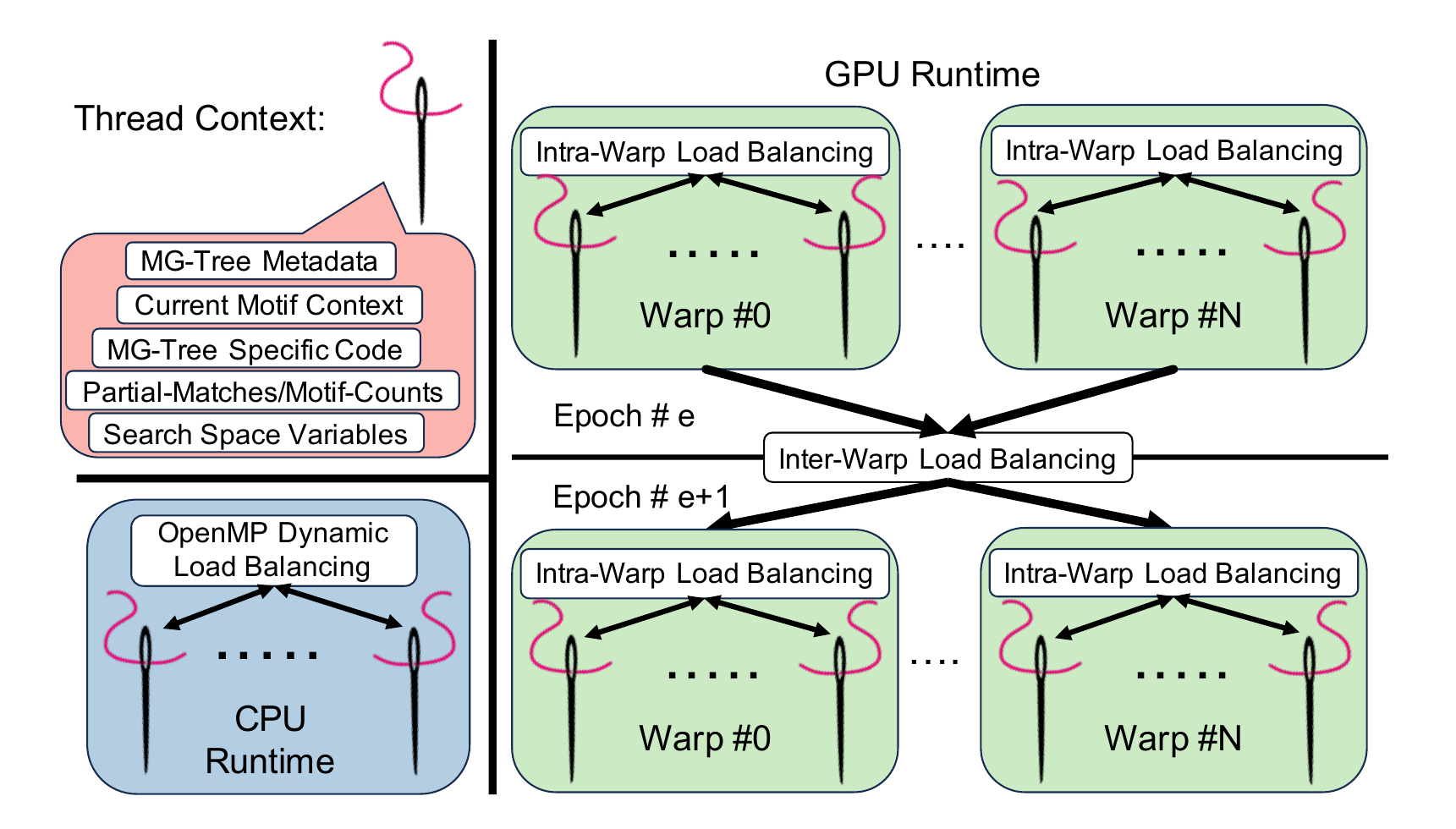}
  \caption{Visualizing Thread Context and Load-Balancing on the CPU and GPU at runtime.}
  \label{fig:runtime}
\end{figure}

The CPU task-manager distributes candidate edges for the first edge across all threads, and utilizes dynamic task scheduling for load balancing.
The GPU task-manager employs a two-level policy, which distributes these candidate edges across warps and then splits the search-tree within a warp~\cite{everest}.
This policy is also used to load-balance across threads within a warp (intra-warp) and the across warps (inter-warp).
Such a multi-granular policy is necessary given the higher thread count and lower computational capability of individual GPU threads compared to CPU threads.
Intra-warp balancing involves periodic polling of idle threads and candidate redistribution using warp-level primitives
Similarly, inter-warp load balancing redistributes the candidates by polling for idle warps across the GPU.
Active threads' search contexts are dumped and then redistributed across all warps on the GPU before resuming mining.
Fig.~\ref{fig:runtime} visualizes the runtime-context held within a thread, the CPU Runtime's load-balancing across CPU threads, and the GPU runtime with its two-tiered balancing 1) within a warp at epoch \#e and \#e+1, and 2) across warps between two epochs.

\begin{figure}
  \centering
  \includegraphics[width=\linewidth]{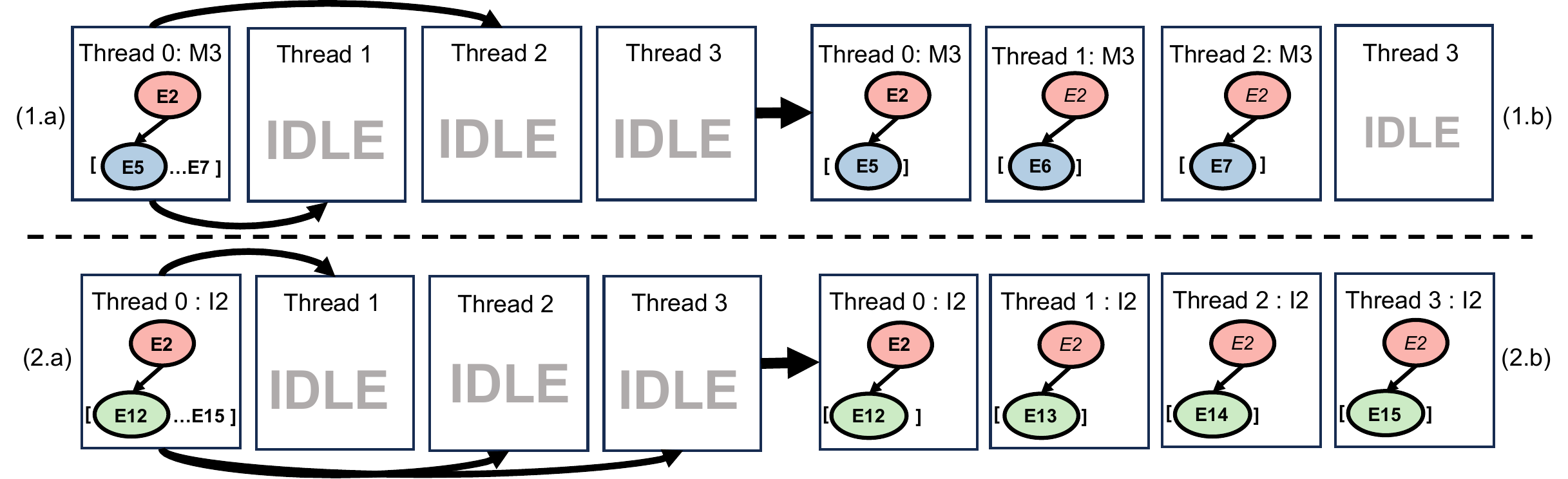}
  \caption{Intra-Warp Load-Balancing distributing candidates across idle warps. Only the source thread (\textit{i.e.,} 0) is allowed to explore sibling Nodes.}
  \label{fig:intra_warp_comining}
\end{figure}

\textbf{Visualizing Intra-Warp Load-balancing:} 
Fig.~\ref{fig:intra_warp_comining} illustrates the intra-warp load balancing mechanism during co-mining the MG-Tree in Fig.~\ref{fig:mg_tree_exmpl_query}, displaying the search-tree and candidates for the second edge in~\textt{I1} and third edge in~\textt{M3} or~\textt{I2}.
Initially, a warp with four threads has only one active thread (Thread 0) processing candidates (\textt{E5}-\textt{E7}) for leaf node M3 (Stage 1.a).
The load balancer redistributes these candidates across threads 0-2 to maximize parallelism (Stage 1.b).
At a later point, when thread 0 runs out of candidates for \textt{M3}, it transitions to mining sibling Node \textt{I2} and generates a new candidate list (\textt{E12}...\textt{E15}).
Unlike thread 0, threads 1 and 2 are restricted from mining any sibling of \textt{M3} (\textit{i.e.,}~\textt{I2}) in order to prevent double-counting of matches, since thread 0 has already started working on candidates for \textt{I2}.
Subsequent load balancing (Stage 2.a) redistributes \textt{I2} candidates from Thread 0 across all four threads, achieving full warp utilization (Stage 2.b)

\begin{figure}
  \centering
  \includegraphics[width=\linewidth]{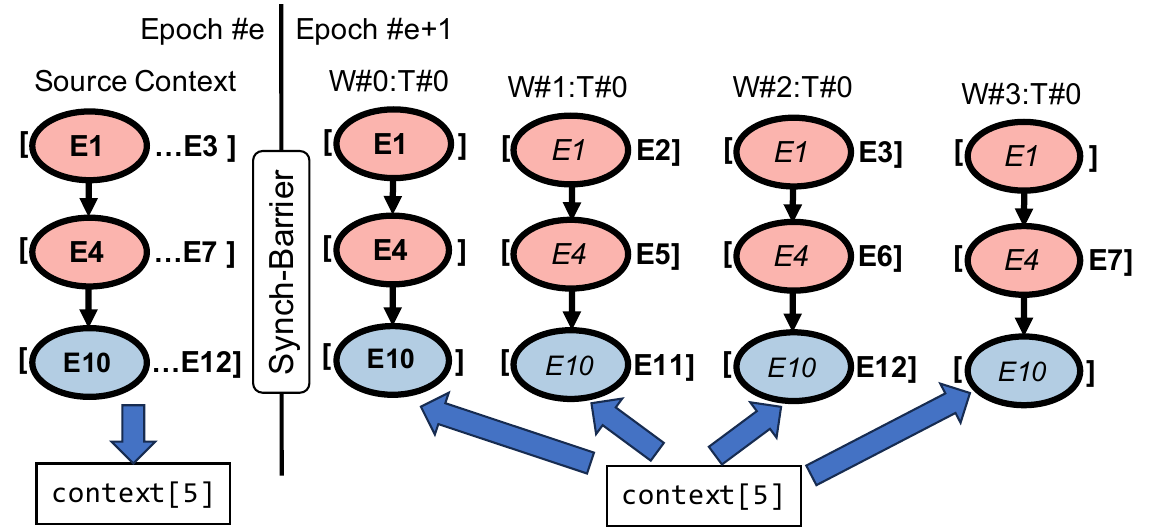}
  \caption{Inter-Warp Load-Balancing distributing candidates across warps from contexts dumped in the previous epoch.}
  \label{fig:inter_warp_comining}
\end{figure}

\textbf{Visualizing Inter-Warp Load Balancing: }
Fig.~\ref{fig:inter_warp_comining} demonstrates the inter-warp load-balancing strategy, the second part of the two-level load-balancing strategy, when mining the MG-Tree in Fig.~\ref{fig:mg_tree_exmpl_query}.
Upon detecting a threshold number of idle warps, the system interrupts the mining process to make all active threads save their context (\textit{i.e.,} search-tree, book-keeping variables \textit{etc.}) and exit.
In Fig.~\ref{fig:intra_warp_comining}, an active thread is saving its context to the fifth position in the \texttt{context} array located in global memory.
This context indicates that the thread is currently exploring candidates for the last edge in \textt{M3} (\textt{E10},\textt{E11},\textt{E12}), with (\textt{E1},\textt{E2},\textt{E3}) and (\textt{E4}-\textt{E7}) as candidates for the first and second edge of \textt{I2} respectively.
Inter-warp load-balancing punctuates periods/epochs of continuous mining with gaps to redistribute the workload.
While the search-tree is duplicated across warps with the same source context, the candidates are partitioned across warps.
It is possible that different threads in the same warp are assigned to work on search-trees distributed from different contexts.
To prevent double counting, the edges that form the search tree are exclusively explored on one thread, with these considered as only a part of the search tree in other threads.
For instance, the threads in warp \#1 and \#2 immediately skip \textt{E10} to process \textt{E11} and \textt{E12} respectively.
In addition to skipping \textt{E10}, the thread in warp \#3 also skips \textt{E4} as a candidate for the second edge in \textt{I2} and moves to considering \textt{E6}.

\textbf{Implementing Load-Balancing:}
\name\ periodically monitors the load-balance rather than continuously calculating the balance factor, owing to the high latency of thread-level synchronization operations.
Intra-warp load balancing (Fig.~\ref{fig:intra_warp_comining}) monitors thread status every (say) INTRA\_INTRVL iterations, with threads voting to redistribute work when idle threads coexist with active ones.
Inter-warp load balancing (Fig.~\ref{fig:inter_warp_comining}) uses global memory to track warp idleness across the GPU every (say) INTER\_INTRVL iterations, triggering redistribution when a threshold of idle warps is detected.
Since global memory access is more expensive than warp-level synchronization, inter-warp monitoring occurs at longer intervals (INTER\_INTRVL > INTRA\_INTRVL), creating the two-tier load balancing strategy visualized in the figures~\ref{fig:intra_warp_comining}~and~\ref{fig:inter_warp_comining}.


\textbf{Cost of Co-mining:}
\name~does not allocate CPU memory co-mining while allocating a relatively small amount of GPU memory (0.1\%-2.5\%) to offload contexts during inter-warp load-balancing.
\name's CPU backend replicates the baseline parallelization scheme and does not incur additional synchronization cost.
However, the GPU backend needs to communicate an additional parameter (compared to the GPU baseline) with other threads in the warp to prevent duplicate exploration of sibling Nodes.
This design decision allows~\textit{\name~ to mine many more motifs simultaneously while incurring little to no memory or synchronization overhead}.

Limiting sibling node exploration to a single thread for correctness underutilizes available parallelism across sibling nodes.
While splitting the candidates and the search tree across threads could mitigate this, it may increase load balancing latency.
Furthermore, GPU threads face constraints like register limits and reduced performance from control-flow divergence.
The additional context and control logic (Algo.~\ref{algo:ctm}) required to enable co-mining thus become an overhead, potentially reducing performance gains.
We can alleviate these drawbacks by minimizing the context and streamlining the control-flow by tuning the source code to the MG-Tree.
\S\ref{sec:opt} discusses optimization strategies to address these challenges.


\section{\name~Design Optimizations}\label{sec:opt}

The co-mining algorithm introduced in \S\ref{sec:design} achieves theoretical efficiency gains by exploiting structural and temporal commonalities across motifs via the MG-Tree.
However, practical implementation on modern hardware architectures requires addressing critical performance bottlenecks unique to CPUs and GPUs.
This section presents a suite of optimizations such as optimized code-generation and load-balancing that bridge the gap between algorithmic innovation and real-world execution efficiency.

\subsection{Motif-Group Specific Code-generation} \label{subsec:opt-codegen}


\subsubsection{CPU Code-generation}\label{subsubsec:cpu-codegen}

For CPU implementations, we generate specialized loops for each level of the MG-Tree search hierarchy.
While the baseline implementation based on prior work~\cite{mackey2018chronological,everest} uses recursive function calls with uniform loop structures, this approach confounds modern branch predictors due to varying iteration ranges across recursion levels.
The code-generation phase for the CPU (Fig.~\ref{fig:exmpl_query},~\circled{4}) unrolls the recursive search into distinct nested loops, each explicitly optimized for its corresponding MG-Tree level.


\subsubsection{GPU Code-generation}\label{subsubsec:gpu-codegen}

\begin{figure*}[]
    \centering
    \includegraphics[width=\textwidth]{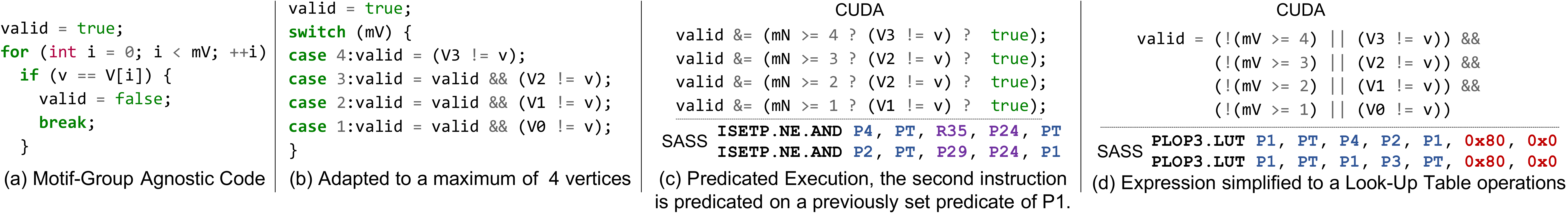}
    \caption{Optimization to reduce warp-divergence and streamline control-flow.}
    \label{fig:gpu_code_opt}
\end{figure*}

GPUs present unique optimization challenges due to their Single Instruction Multiple Thread (SIMD) architecture and constrained register resources.
Unlike CPUs that excel at handling complex control flow through speculative execution and sophisticated branch predictors, GPUs require fundamentally different optimizations to avoid performance pitfalls like warp divergence (threads in a warp executing different code paths) and register pressure (exceeding limited per-thread register capacity).
Our GPU code generation strategy employs three synergistic optimizations to address these challenges while maintaining the algorithmic benefits of co-mining.


\textbf{Register-Bound Context Mapping.} 
Fig.~\ref{fig:gpu_code_opt}(a) illustrates a portion of the code that enforces structural constraints by ensuring that the new candidate vertex,~\texttt{V}, has not been matched before by comparing ~\texttt{V} with all vertices upto \texttt{V[mV-1]}, where~\texttt{mV}~is the number vertices that have been mapped and~\texttt{V}~is the array of vertices.
We store these values (and the most of the context) in thread-local memory since the code has to flexibly work with any motif group with different number of vertices, edges, and motifs.
This flexibility results in costing latency since the GPU incurs a memory operation to the shared or global memory.
Given the MG-tree, we can replace the dynamic-array based context with fixed registers (\textit{e.g.,} \texttt{V0}-\texttt{V3}) like in Fig.~\ref{fig:gpu_code_opt}(b), effectively hard-coding the mapping state for known motif sizes and eliminating memory accesses to the context.

In GPU architectures, warps execute instructions in lockstep across all 32 threads.
Divergent control flow serializes execution, forcing subsets of threads to wait at synchronization points until all warp lanes complete their current path.
The synchronization overhead, of tracking divergent paths and maintaining thread masks, incurs substantial latency, leading to a phenomenon known as warp divergence.
While replacing loops with the switch-case in Fig.~\ref{fig:gpu_code_opt}(b) reduces total branches, divergence persists when threads in a warp have different values for~\textt{mV}.

\textbf{Predicated Control-Flow.}
Modern GPUs support predicated execution, where instruction execution is conditional on a Boolean value stored in predicate registers.
This mechanism enables thread-specific control flow without explicit branching—instructions execute as no-ops (NOPs) when their associated predicate evaluates to false, thus eliminating warp-divergence and streamlining control-flow.
As shown in Fig.~\ref{fig:gpu_code_opt}(c), we utlize this feature to predicate the structural constraint checks based on the value of \texttt{mV}.
The figure also contains two examples of predicated instructions in NVIDIA SASS, a low-level assembly language for NVIDIA GPUs.
The first instruction is unconditionally executed since it is predicated on the always-true predicate \texttt{PT}, and the second one is dependent on a previously set predicate register \texttt{P1}.
While predication avoids branch divergence, its application is limited by two constraints: 1) only arithmetic/logic operations can be predicated, and 2) predication is more effective with short code blocks, since longer predicated blocks incur higher latencies by issuing instructions from both the \texttt{if-then} and \texttt{else} parts even when the threads do not diverge.

\textbf{Expression Simplification.}
Rewriting multiple statements in Fig.~\ref{fig:gpu_code_opt}(c) as a single boolean expression exposes further opportunities for the compiler.
The predicated logic to test a specific vertex represents an implication from the value of \texttt{mV} to a check on the vertex, \textit{i.e.,} $A\rightarrow B$, and can simplified to a boolean expression resembling $\neg A \lor B$.
These transformations enables the compiler to fuse multiple logical operations into 8-bit LookUp-Table (LUT) instructions, as illustrated in Fig.~\ref{fig:gpu_code_opt}(d).
This reduces the instruction count since multiple operations were replaced by a single LUT instruction and also reduces register pressure since registers are no longer required to hold as many intermediate values of an expression as before, this improving occupancy.

In summary, by taking advantage of the compile-time knowledge of the maximum context size from the MG-Tree, and strategically applying optimizations across the code base, \name~mitigates warp divergence, alleviates register pressure, and minimizes instruction counts. This synergy between algorithmic design and hardware awareness enables a more efficient exploitation of computational resources, yielding significant performance improvements.

\subsection{GPU-Specific Load-balancing} \label{subsec:gpu-load-balance}

While the two-tier load-balancing scheme effectively exploits parallelism across search-tree candidates, it under-utilizes parallelism available across Nodes in the MG-Tree.
We address this limitation using two complementary optimizations: \textit{sibling-splitting} for intra-warp parallelism and \textit{multi-offloading} for inter-warp parallelism.

\begin{figure}[]
    \centering
    \includegraphics[width=\linewidth]{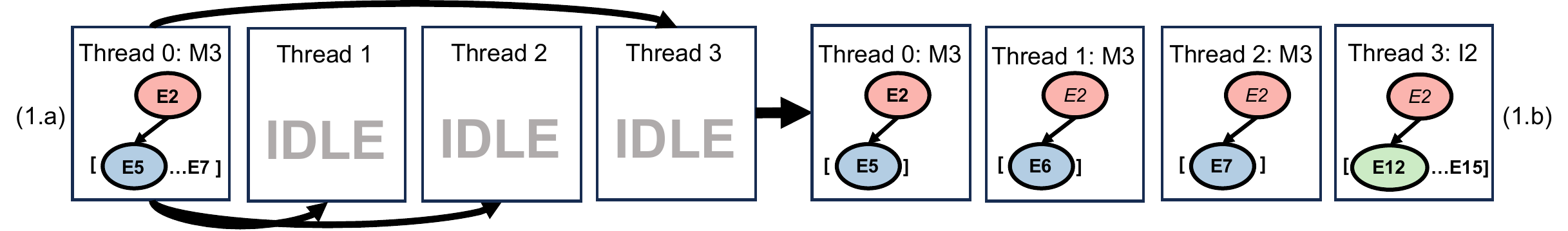}
    \caption{Sibling-Splitting for Intra-Warp Load-balancing.}
    \label{fig:intra_warp_opt}
\end{figure}

\textbf{Sibling-Splitting for Intra-Warp Parallelism.}
Fig.~\ref{fig:intra_warp_opt} demonstrates the sibling-splitting optimization during intra-warp load balancing.
The detection of idle threads triggers the load-balancing similar to \S\ref{subsec:runtime} (Stage 1.a). 
For all the active threads, the system checks whether the Node being mined has an unexplored sibling according to the order determined in Algo.~\ref{algo:mg_tree_const}.
For thread 0 mining \textt{M3} in Fig.~\ref{fig:intra_warp_opt}, the unexplored sibling would be \textt{I2}.
Threads with such sibling Nodes 1) nominate an idle thread participating in the to mine the sibling (\textt{I2} -> thread 3), and 2) distributes the candidates of its search-tree across other threads (threads 0 - 2).
Thread 3 then obtains the candidate list for \textt{I2} while retaining the search-tree for \textt{I2}'s parent Node, \textt{I1}.
This approach enables the exploration of sibling Nodes in parallel, increasing the availability of candidates to reduce the number of idle threads.

\begin{figure}[]
    \centering
    \includegraphics[width=\linewidth]{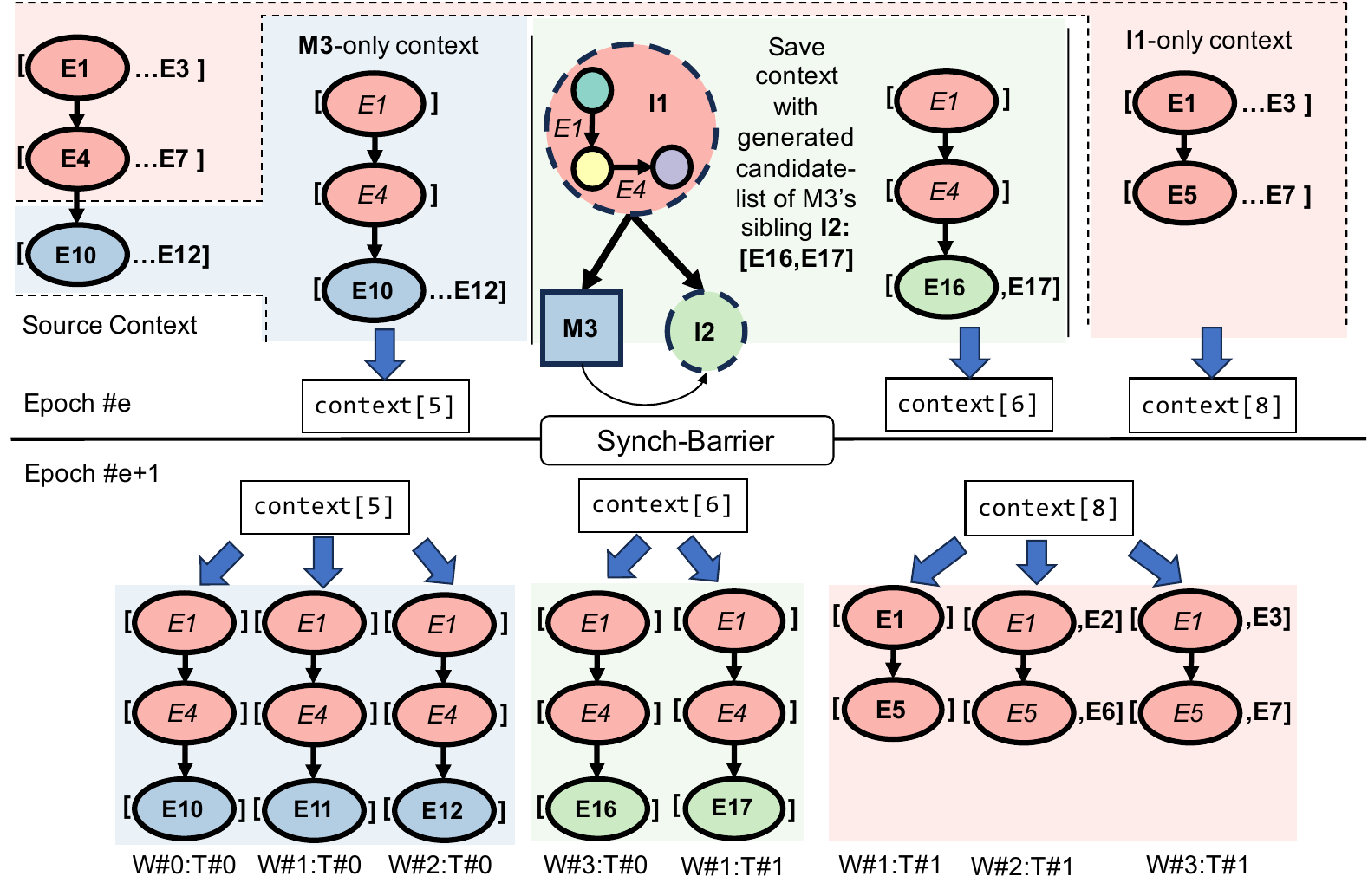}
    \caption{Optimization for Inter-Warp Load-balancing.}
    \label{fig:inter_warp_opt}
\end{figure}

\textbf{Multi-Offload for Inter-Warp Parallelism}
is an optimization enabling concurrent exploration of Nodes across the MG-Tree hierarchy by decomposing the search context across multiple levels.
Fig.~\ref{fig:inter_warp_opt} illustrates this approach during inter-warp load balancing, where a thread was mining for M3 with candidates for both edges in parent I1 before being interrupted to offload its context.
Since siblings can be mined independent of each other, the optimization exposes this parallelism by creating separate contexts for candidates of each sibling (context[5] for M3, context[6] for I2) while preserving the search tree of their parent (I1).
Since siblings are not part of the original search-tree, the candidate lists for a sibling is generated before its context is offloaded into global memory (\textit{e.g.,} for I2 in Fig.~\ref{fig:inter_warp_opt}).
The same process is repeated for the parent Node and its siblings until the root Node is reached, with the search tree being trimmed to reflect the shallower depth in the MG-tree.

In the case of Fig.~\ref{fig:inter_warp_opt}, a context is created that only contains I1's candidates (saved in context[8]).
Note that all child Nodes have been explored with E4 a candidate for the second edge in I1, the I1-only context skips E4 and moves onto E5.
This hierarchical decomposition exposes parallelism across as many levels as possible in the MG-tree with a given context by offloading multiple contexts for siblings in each level.
When resuming mining in the next epoch in Fig.~\ref{fig:inter_warp_opt}, the decomposed contexts are able to keep eight threads busy instead of just four threads without the optimization (Fig.~\ref{fig:inter_warp_comining}).
Observe that Sibling-Splitting paired with inter-warp load-balancing has a similar effect to that of Multi-Offload: by starting the search for siblings at an early stage, sibling-splitting creates multiple contexts from a single context, albeit constrained to the same level, and these multiple contexts are offloaded during load-balancing.

\textbf{Additional Resource Footprint.}
While the above optimizations expose additional parallelism in the workload, they can introduce trade-offs that must be carefully managed.
As intra-warp load balancing is invoked frequently, we constrain sibling-splitting to explore only one sibling at a time to minimize the overhead of exploring multiple siblings.
The multi-offload strategy, while effective in distributing work, can incur additional instructions to offload multiple contexts.
Our experiments reveal that these optimizations incur negligible overheads: dynamic instruction counts increase by $\le$6\% and occupancy is reduced by $\le$1\%.
These results confirm the practicality of our approach, as the performance benefits of enhanced parallelism outweigh the modest resource costs (\S\ref{sec:eval-results}).

\section{Evaluation Methodology}\label{sec:eval-method}

\begin{figure*}[]
    \centering
    \includegraphics[width=\textwidth]{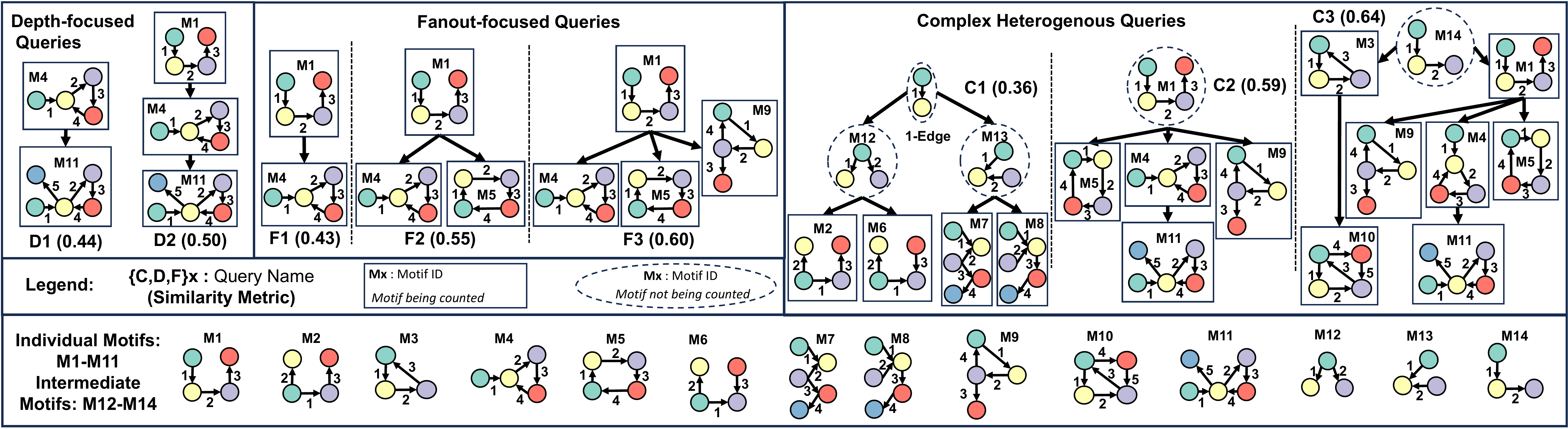}
    \caption{MG-Trees of Motif Groups, with respective (SM).}
    \label{fig:mg_tree_eval}
\end{figure*}
\textbf{The Baselines} compared against \name's CPU and GPU implementations respectively are the work proposed by Mackey et.al.~\cite{mackey2018chronological} and Everest~\cite{everest}.
Both baseline methods exploit intra-query parallelism, by spreading out the search space for a single query across multiple threads.

\noindent
\textbf{The Hardware Setup} consists of a server with an Intel Xeon Platinum 8380 CPU (40 cores, 80 threads) with 1TB of main memory, and an NVIDIA A40 GPU with 48GB GDDR6 memory.


\noindent
\textbf{Five Datasets} of real-world temporal graphs spanning social networks, blockchain transactions, and internet traffic (Table~\ref{table:input_dataset}) were used to evaluate \name.
Since the \textt{equinix (eqx)} dataset captures the exchange of internet packets between computers of two cities, making it a bipartite graph.
Due to memory capacity limitations of our GPU, we subsample the massive \textt{eqx} dataset to 37.5\% of its original edges while preserving temporal characteristics.

\noindent
\textbf{Eight Queries} were created by combining fourteen motifs, of which \texttt{M1}~-~\texttt{M11} are counted and \texttt{M12}~-~\texttt{M14} are only intermediates, which were used in prior work~\cite{everest, mackey2018chronological,sarpe2021presto,Kosyfaki2018FlowMI}, and are illustrated in Fig.~\ref{fig:mg_tree_eval}.
The queries cover three categories:
\begin{description}
    \item[Depth-focused:] Deepening MG-Trees (\textt{D1}-\textt{D2}).
    \item[Fanout-focused:] Widening MG-Trees (\textt{F1}-\textt{F3}).
    \item[Complex Heterogenous:] Variety of sizes and overlap~(\textt{C1}-\textt{C3}).
\end{description}

To capture the notion of similarity among motifs, we define the Similarity Metric (SM) for a motif group MG and its MG-Tree as,
$$
SM(MG, MG\text{-}Tree) = 1 - \frac{\sum_{M \in MG\text{-}Tree} (||E_M|| - ||E_{M.parent}||)}{\sum_{M \in MG} ||E_M||}
$$
where $||E_M||$ is the number of edges in motif $M$.
The denominator represents the aggregate edge count across all motifs in $MG$, while the numerator captures the cumulative incremental edge count relative to their parent Nodes in the $MG\text{-}Tree$.
Higher inter-motif similarity reduces parent-child edge differentials, thereby decreasing the numerator and increasing SM values.
Motif groups with elevated SM scores typically exhibit greater opportunities for computational reuse through our MG-Tree traversal.
However, realized speedups remain contingent on system-specific factors including hardware utilization and load-balancing efficiency.
Since the timescale of events varies across the datasets, we employ specific $\delta$ values that reflect meaningful time-windows and limit run-time~\cite{everest}.

\begin{table}
  \centering
  \scriptsize 
  \begin{tabular}{ c | c | c | c | c | c}
  \hline
  \rule{0pt}{5pt}
  \textbf{Graph} & \textbf{\#Vertices} & \textbf{\#Temporal} & \textbf{\# Static}        & \textbf{Time} & \textbf{$\delta$}
  \\             &                     & \textbf{Edges}      & \textbf{Edges}            & \textbf{Span} & \textbf{Window}
  \tabularnewline
  \hline
  \hline
  wiki-talk (wtt)~\cite{snap} & 1,140,149 & 7,833,140 & 2,787,968 & 6.24 years & 1 day\\
  stackoverflow (sxo)~\cite{snap} & 2,601,977 & 63,497,050 & 34,875,685 & 7.6 years & 1 day\\
  reddit-reply (trr)~\cite{liu2018sampling} & 8,901,033 & 646,044,687 & 435,290,421 & 10.1 years & 10 h\\
  ethereum (eth)~\cite{ethdatasets} & 66,323,478 & 628,810,973 & 186,064,655 & 3.58 years & 1 h\\
  equinix (eqx)~\cite{sarpe2021presto} & 6,208,412 & 872,124,829 & 29,766,272 & 23.46 mins & 3.6 ms\\
  \hline
  \end{tabular}
  \caption{Temporal graph datasets used for evaluation.}
  \label{table:input_dataset}
\end{table}

\section{Evaluation Results}\label{sec:eval-results}

%

\begin{figure*}
    \centering
    \begin{minipage}{0.39\linewidth}
        \centering
        \includegraphics[width=\textwidth]{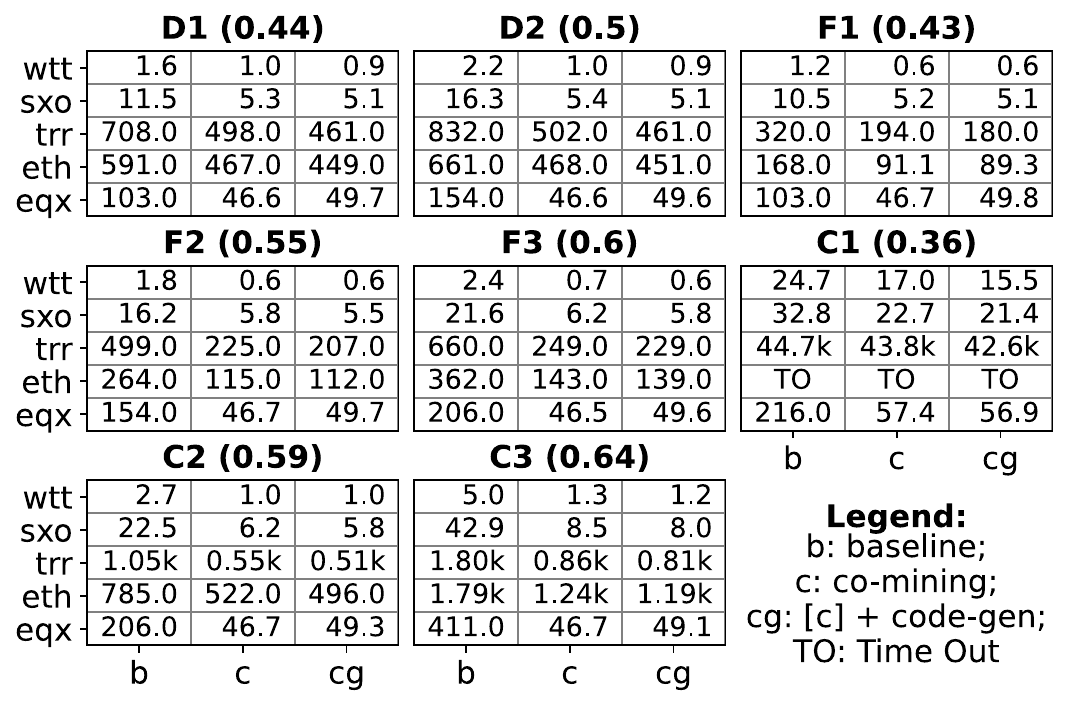}
        \caption{CPU Timings (seconds).}
        \label{fig:cpu-timings}
    \end{minipage}\hfill%
    \begin{minipage}{0.61\linewidth}
        \centering
        \includegraphics[width=\textwidth]{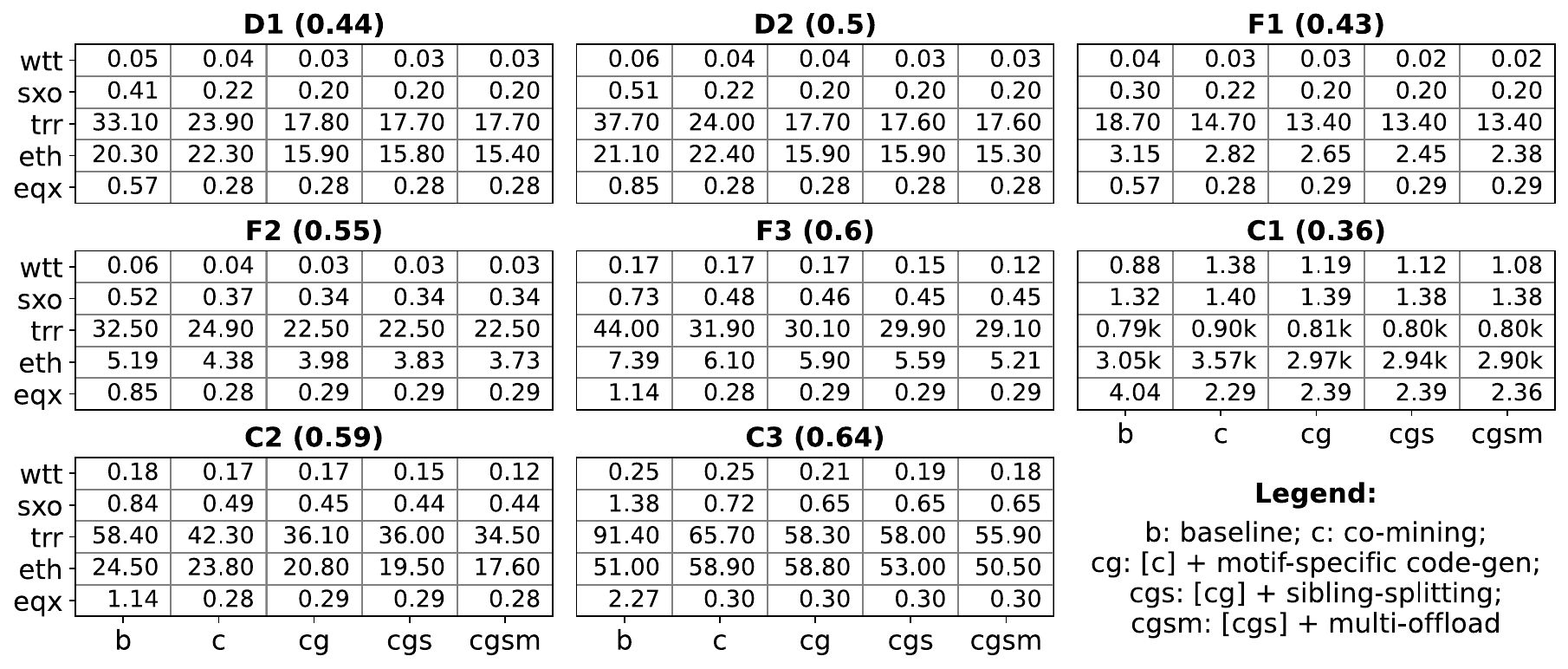} 
        \caption{GPU Timings (seconds).}
        \label{fig:gpu-timings}
    \end{minipage}
\end{figure*}

\begin{figure*}
    \centering
    \begin{minipage}{0.49\linewidth}
        \centering
        \includegraphics[width=\textwidth]{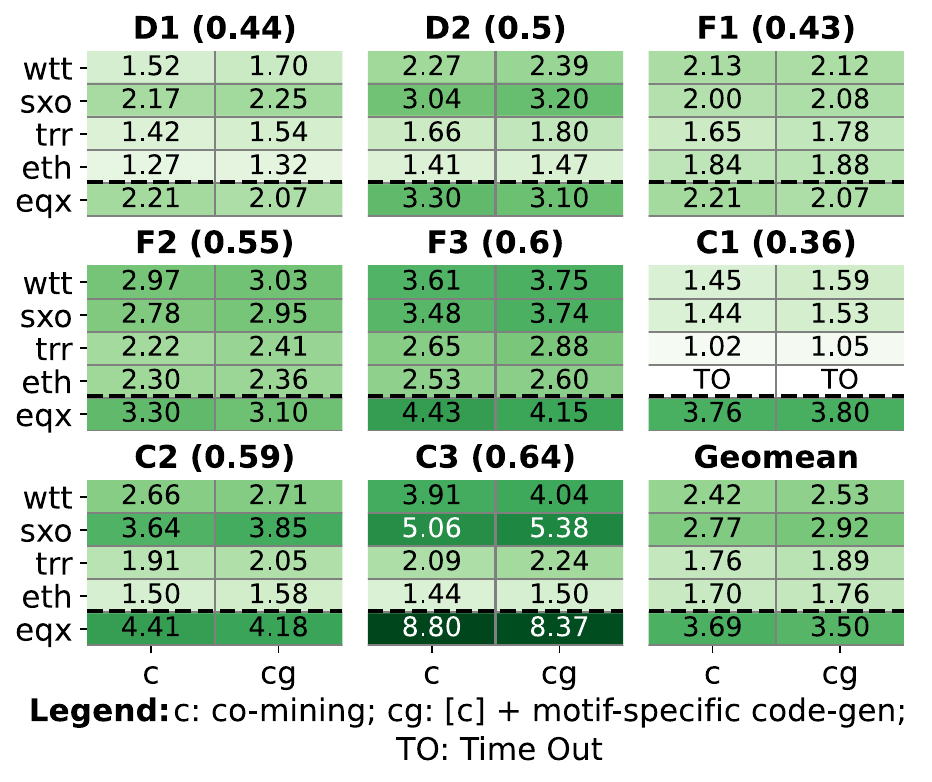}
        \caption{Breakdown of performance improvements of different optimizations on the CPU, compared to the baseline.}
        \label{fig:cpu-speedups}
    \end{minipage}\hfill%
    \begin{minipage}{0.49\linewidth}
        \centering
        \includegraphics[width=\textwidth]{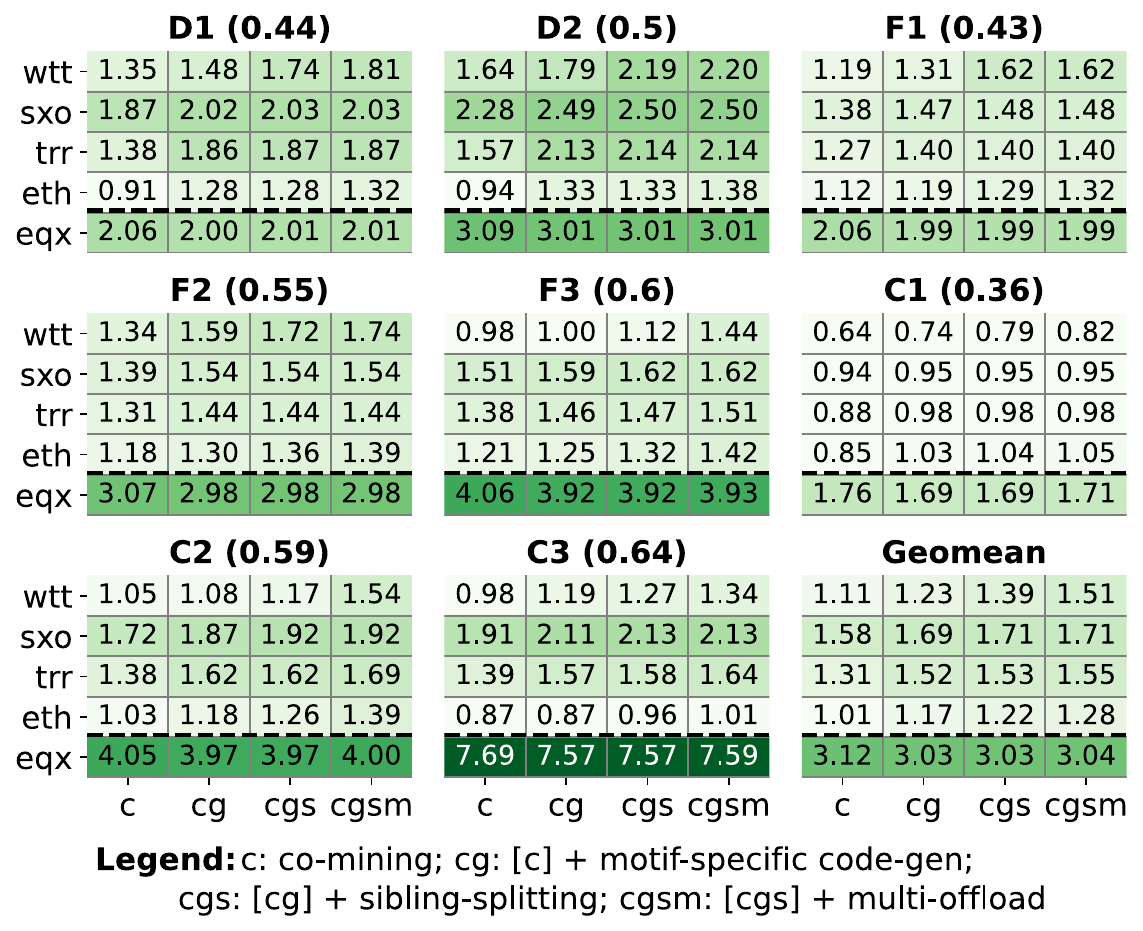} 
        \caption{Breakdown of performance improvements of different optimizations on the GPU, compared to the baseline.}
        \label{fig:gpu-speedups}
    \end{minipage}
\end{figure*}

\textbf{High Level Summary: }
The evaluation results reveal substantial performance improvements from integrating co-mining techniques with our optimizations.
Figures ~\ref{fig:cpu-timings} and ~\ref{fig:gpu-timings} capture the timings, in seconds, on the CPU and GPU respectively.
For each query (title), the baseline and a set of co-mining optimizations (columns) are tested across multiple datasets (rows).
Figures ~\ref{fig:cpu-speedups} and ~\ref{fig:gpu-speedups} capture the individual speedups for each query over the baseline for the corresponding  set of optimizations on the CPU and GPU respectively, with the "Geomean" numbers being the geometric mean of the speedup across all queries for a particular dataset.
On the CPU (Fig.~\ref{fig:cpu-speedups}), individual speedups range from 1.05$\times$ to 8.37$\times$, with co-mining alone yielding an overall average improvement of 2.35$\times$ and code-generation pushing it up to an average of 2.46$\times$.
On the GPU (Fig.~\ref{fig:gpu-speedups}), individual speedups range from 0.82$\times$ to 7.59$\times$, just using co-mining yields an overall average improvement of 1.48$\times$ with other optimizations raising it to 1.73$\times$.
While the absolute time saving for the experiments on the GPU, spanning shorter runtimes (milliseconds to minutes), may appear modest at best, the speedups remain practically significant in high-throughput analytical environments where thousands of such queries are executed daily~\cite{jindal2018selecting,wang2023real,ma2018query,van2024tpc}.
By combining the multi-query processing approach with the intra-query parallelism implemented in the baselines, the overall process is more efficient when exploring the search space in parallel.
These gains are strongly influenced by the degree of structural similarity among motifs: motif-groups with higher overlap benefit more from the co-mining strategy, whereas groups with minimal overlap (e.g., the \textt{C1} motif group) exhibit limited improvements and, in some cases, even a performance degradation on the GPU due to increased resource constraints.
Note that mining \textt{C1} on the \textt{eth} dataset on the CPU exceeded the time-limit of 24 hours and has been omitted from our comparisons.

\textbf{\name~effectively exploits structural similarities among motifs.}
\textt{D2} experiences higher speedups than \textt{D1} due to the implicit mining of \textt{M1} before mining \textt{M4}, even if \textt{M1} is not counted explicitly.
This trend is consistently observed when moving from \textt{F1} to \textt{F3} and \textt{C2} to \textt{C3}, where the expansion of the MG-tree to include more structurally similar motifs correlates with improved speedups. 
\textt{C1}, characterized by low inter-motif overlap (S.M.), proves challenging to optimize, resulting in the lowest speedups and even performance degradation on the GPU due to reduced occupancy and increased warp divergence compared to the baseline.

\textbf{Dataset characteristics} significantly influence performance gains. 
On the CPU, all datasets benefit from co-mining and most of them benefit from the code-generation as well.
The bipartite \textt{eqx} dataset exhibits exceptionally high speedups on both CPU and GPU platforms, due to the fact that bipartite graph cannot allow motifs that connect vertices in the same disjoint partition, naturally eliminating any match.
Consider \textt{D1}, where  \textt{M1} can be found in a bipartite graph since (\img{figures/node_a.pdf}, \img{figures/node_c.pdf}) could $\in$ partition \#0 and  (\img{figures/node_b.pdf}, \img{figures/node_d.pdf}) $\in$ partition \#1. 
Proceeding to \textt{M4} after matching \textt{M1}, we would fail to find any matches since there are no edges between \img{figures/node_b.pdf} and \img{figures/node_d.pdf} as they $\in$  partition \#1.
This in-turn prunes the search for any descendents of \textt{M4}, \textit{i.e.,} \textt{M11}, since they depend on matching \textt{M4} first.
This way, the algorithm is able to eliminate multiple redundant searches required for \textt{M4} and \textt{M11} when mining them individually.
\textt{wtt}, \textt{trr}, and \textt{eth} datasets benefit from all GPU optimizations, although the extent varies by motif-group.
\textt{sxo}'s performance peaks at the \textt{cgs} optimization, suggesting a well-balanced workload distribution that obviates the need for inter-warp load balancing.
The reduced speedups observed for \textt{eth} and \textt{trr} across all queries stems from their elevated motif match density (\textit{i.e.,} $\sigma = ||matches|| \div ||E_G||$).
With a higher $\sigma$, threads process significantly larger candidate sets per edge, prolonging exploration of individual MG-Tree Nodes and delaying transitions to sibling Nodes.
This increases the serialization of the search across the MG-tree hierarchy, reducing the efficacy of co-mining.
Conversely, \textt{eqx} achieves superior performance due to its bipartite structure leading to the smallest $\sigma$ among all datasets, enabling aggressive pruning of the search space.



\begin{figure}
    \centering
    \begin{subfigure}[]{\linewidth}
        \centering
        \includegraphics[width=\textwidth]{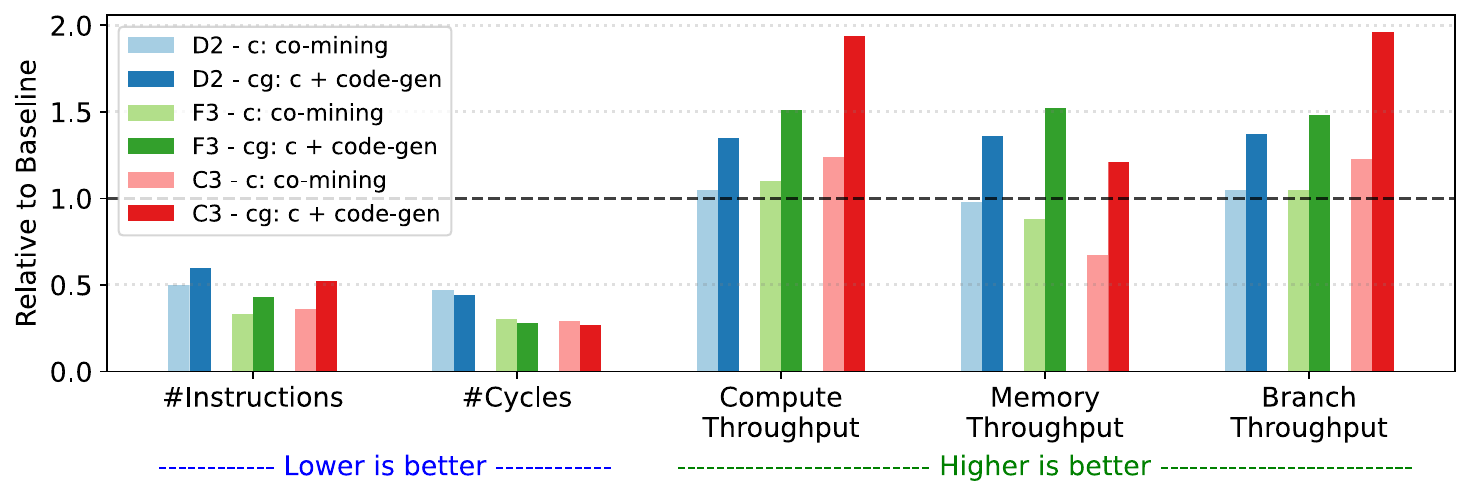}
        \caption{CPU Backend}
        \label{fig:cpu-perf-breakdown}
    \end{subfigure}
    \vfill
    \begin{subfigure}[]{\linewidth}
        \centering
        \includegraphics[width=\textwidth]{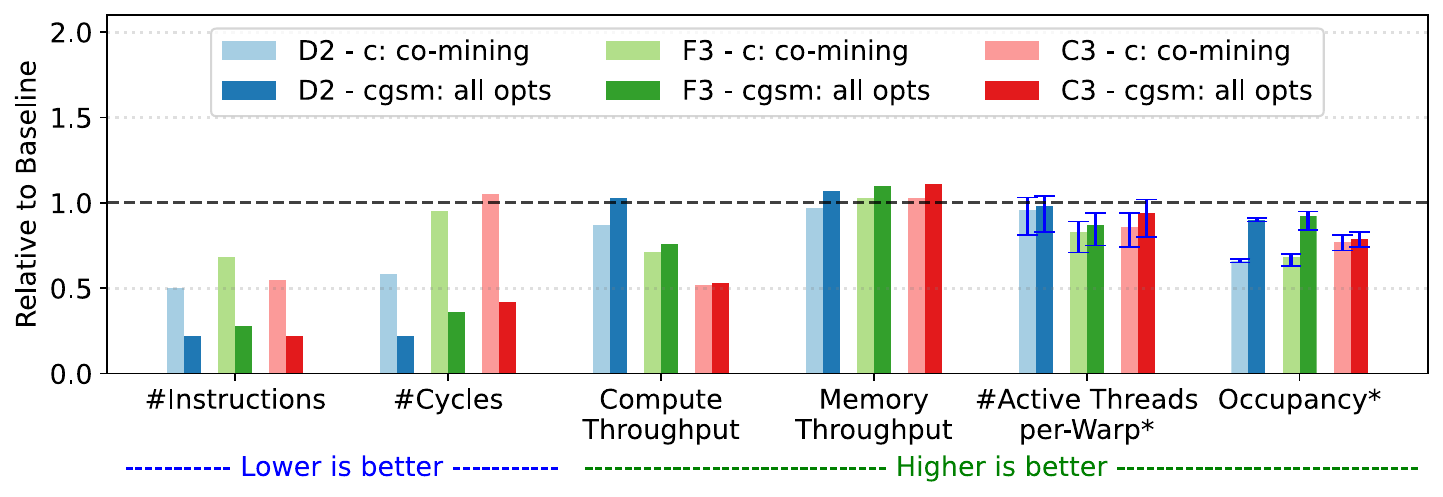}
        \caption{GPU Backend}
        \label{fig:gpu-perf-breakdown}
    \end{subfigure}
    \caption{Architectural Metrics for mining \textt{D2},~\textt{F3} and \textt{C3} on~\textt{wtt}, relative to the baseline.}
    \label{fig:perf-breakdown}
\end{figure}


\textbf{Performance Analysis: }
To explain the underlying factors contributing to these performance improvements, we analyzed key architectural metrics for both CPU (Fig.~\ref{fig:cpu-perf-breakdown}) and GPU (Fig.~\ref{fig:gpu-perf-breakdown}) implementations, which have been collected by running the queries for motif groups \textt{D2}, \textt{F3} and \textt{C3} on \textt{wtt}.
The figures compare five architectural metrics of performing only co-mining and co-mining with optimizations, with the baselines of the respective implementation.
The "\#Active Threads per Warp" and "Occupancy" metrics for the GPU baseline were computed as time-weighted averages across the individual baseline kernels.
This approach accounts for the varying occupancy characteristics and performance profiles of the baseline kernels.
The error bars indicate the extent to which the optimized kernels outperform / underperform the baseline kernels.
Using standalone CPU co-mining or combining it with code-generation effectively reduces the number of instructions executed, an indication that the system overall is performing less work.
S
The efficacy of code-generation to address branch prediction is reflected in the higher branch-instruction throughput (\ref{fig:cpu-perf-breakdown}).
While GPU co-mining reduces instruction counts, its efficacy diminishes with larger motif groups due to reduced occupancy and increased warp divergence.
Our optimizations, aimed at streamlining control flow and eliminating unnecessary instructions, partially mitigate these issues by increasing the average number of active threads per warp.
However, the complex control flow inherent in co-mining constrains the overall effectiveness of multi-offload and sibling-splitting optimizations.
We also performed an experiment to evaluate the efficacy of improving the occupancy at the cost of increased memory operations.
We chose \textt{C3} motif group since its kernel exhibits the lowest occupancy (44\%) among all motif groups due to high register usage for maintaining motif counts.
Offloading counters to thread-local memory increased occupancy to 70\% but yielded only marginal performance improvements due to increased memory accesses, underscoring the trade-off between occupancy and memory access efficiency in GPU implementations.

\textbf{Memory Footprint:}
Although we do not allocate any extra global memory explicitly for the CPU implementation, the additional footprint has ranged between 100KB less to 180KB over the footprint of CPU baseline mining only one motif.
We dismiss this as noise, especially when the smallest dataset has a footprint of 3GB in CPU RAM.
That said, the GPU backend does allocate extra global memory to enable the inter-warp load-balancing.
This is due to the extra context needed to guide the search in the case of co-mining, as opposed to single motif mining.
This extra space turns out to be entirely dependent on the motif group: 14MB for F3, 16MB for D2, 20MB for D3.
Each motif group has different requirements since it's thread-local context would scale with longer motifs or wider MG-Tree.
These overheads are still relatively small when comparing the smallest of wtt at 800MB (2.5\% overhead) to the the largest dataset of trr at 17Gb (0.1\%).
Also, the GPU baseline and \name's GPU backend use a more space efficient graph representation for the GPU (800MB for wtt) than CPU (3GB for wtt) to mitigate the space constraints of GPUs.
This design decision allows~\name~to process many more motifs concurrently while maintaining a similar footprint.

\textbf{GPU Footprint:}
The larger context held within a GPU thread to enable co-mining reduces the amount of parallelism available to exploit, with the limited register file acting as a bottleneck on the number of active threads/blocks.
As shown in the Table~\ref{tab:comining-gpu-impact}, this increased register requirement reduces the number of active GPU blocks, effectively reducing the number of threads that can execute simultaneously, and creating a performance trade-off as we scale the number of co-mined motifs.

\begin{table}[h]
\begin{tabular}{|r|r|r|r|}
\hline
\multicolumn{1}{|l|}{\textbf{\#Motifs}} &
  \multicolumn{1}{l|}{\textbf{\#Registers}} &
  \multicolumn{1}{l|}{\textbf{\#Blocks}} &
  \multicolumn{1}{l|}{\textbf{Reduction in \#Blocks}} \\ \hline
1 & 44 & 1092 & 0\%  \\ \hline
4 & 55 & 1008 & 8\%  \\ \hline
8 & 67 & 756  & 31\% \\ \hline
\end{tabular}
\caption{
Impact of co-mining on GPU register utilization and thread occupancy
}
\label{tab:comining-gpu-impact}
\end{table}

\begin{figure}
    \centering
    \begin{subfigure}[b]{\linewidth}
        \centering
        \includegraphics[width=\textwidth]{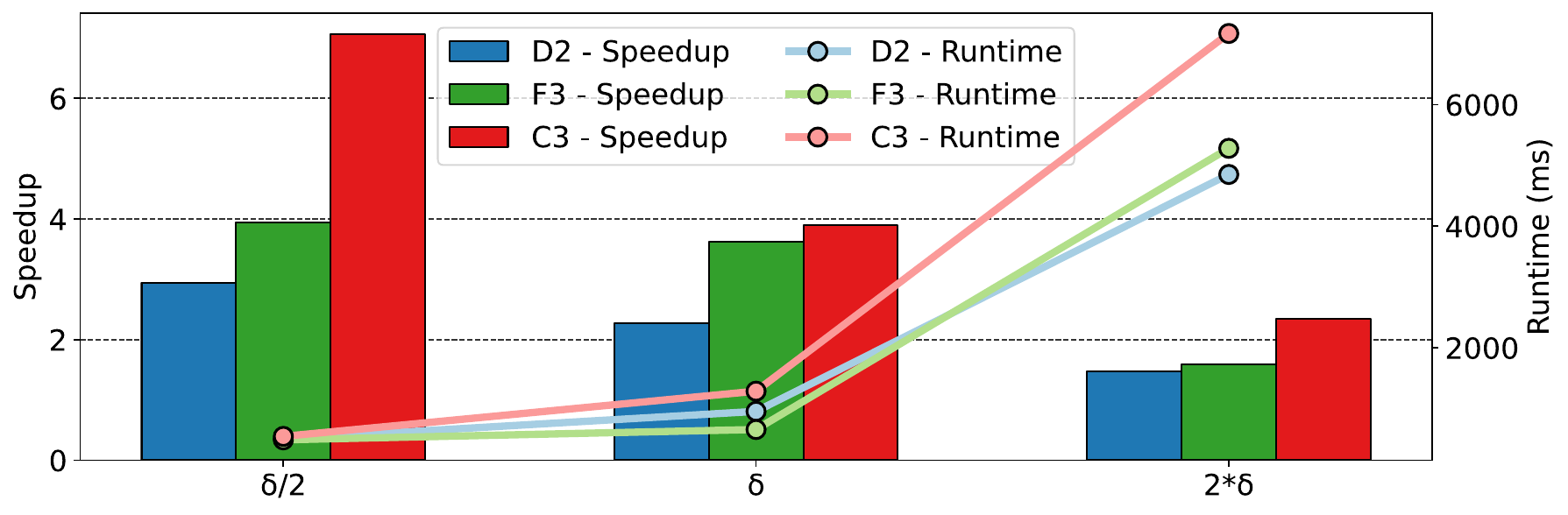}
        \caption{CPU Backend}
        \label{fig:cpu-delta-ablation-study}
    \end{subfigure}
    \vfill
    \begin{subfigure}[b]{\linewidth}
        \centering
        \includegraphics[width=\textwidth]{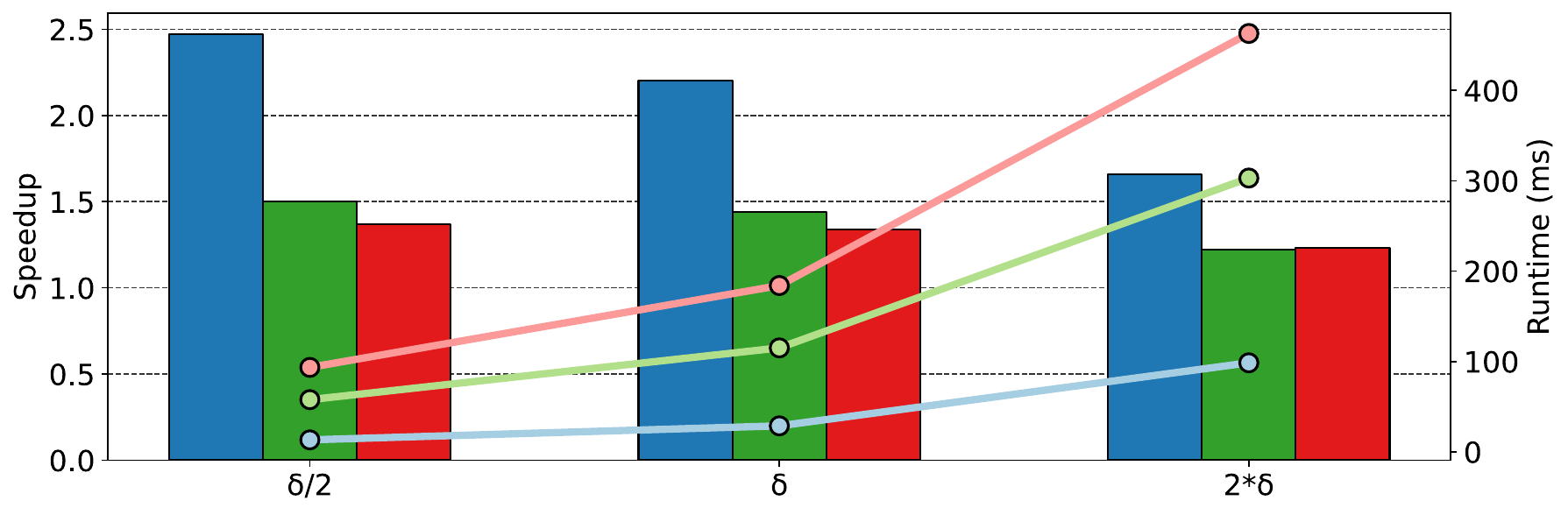}
        \caption{GPU Backend (same legend as (a))}
        \label{fig:gpu-delta-ablation-study}
    \end{subfigure}
    \caption{Effect of scaling $\delta$ on Speedup and Runtime.}
    \label{fig:delta-study}
\end{figure}

\textbf{Effect of $\delta$:}
We evaluated the efficacy of \name~under varying temporal constraints, by comparing its performance to the baseline, mining \textt{D2}, \textt{F3}, and \textt{C3} on the \textt{wtt} dataset, with time-window settings of $\delta/2$, $\delta$, and $2*\delta$.
The results in Figs.~\ref{fig:cpu-delta-ablation-study} and ~\ref{fig:gpu-delta-ablation-study} indicate that shorter time-windows lead to a greater speedup relative to the baseline on both the CPU and GPU.
Longer time windows expand the candidate set and widen the search tree, potentially across multiple levels.
Such wide search trees lead to load-imbalances in the system when they are not split-up across multiple compute units.
This is particularly evident with the CPU backend since it performs balancing only at the top-level.
Whereas the GPU implementation, which employs finer-grained load-balancing across all levels in the search-tree, is less adversely affected.

\textbf{Heuristic for Co-Mining:}
The efficacy of co-mining for a given motif group and dataset combination can be anticipated by analyzing the structure of the graph, S.M and $\delta$.
Co-mining on a bipartite graph has always resulted in a performance improvement on both platforms, as it disallows motif matches to have edges incident within same partition.
For co-mining to be more efficient than the GPU baseline, it needs a minimum S.M. to offset the tighter architectural constraints (Tab.~\ref{tab:comining-gpu-impact}), which we found to be 0.44 from our evaluation.
A user with flexibility to choose $\delta$ could reduce the time-window to improve performance (Fig.~\ref{fig:delta-study}).
Based on our evaluation results, we propose a heuristic (Lst.~\ref{lst:heuristic}) that determines whether co-mining would be beneficial.

\begin{figure}[]
\centering
\begin{lstlisting}[style=mystyle,language=Python, caption=Heuristic for Co-Mining, label=lst:heuristic]
if is_bipartite(data_graph):
    co_mining = True
elif platform == "GPU" and SM < 0.44:
    co_mining = False
else:
    choose_smaller(delta)
    co_mining = True
\end{lstlisting}
\end{figure}



\section{Related Work} \label{sec:related-work}


\noindent
\textbf{Multi-Query Optimization (MQO)}: 
\name~addresses the unique challenges of MQO in the context of temporal motif mining.
In the past, GEqO \cite{haynes2023geqo} pioneered ML-based identification of semantically equivalent subexpressions, while Ma \textit{et al.}~\cite{ma2024efficient} extended MQO to continuous subgraph matching in dynamic graphs.
MapReduce adaptations~\cite{wang2013multi} demonstrated MQO's versatility across paradigms.
While these approaches focus on relational queries or static graph processing,~\name~extends the concept of multi-query optimization to the domain of temporal graph mining, identifying structural and temporal commonalities across multiple motifs.

\noindent
\textbf{Static Graph Mining}
systems count matches of a pattern based on the structure of the query pattern~\cite{peregrine,peregrineFollowup,graphzero,automine,graphpi,stmatch22, G2Miner2022OSDI,rapidmatch,sun2023efficient}.
Notable contributions include Arabesque~\cite{arabesque}, which introduced a distributed framework for graph mining, and Peregrine~\cite{peregrine,peregrineFollowup}, which optimized pattern-aware exploration.
G2Miner~\cite{G2Miner2022OSDI} further advanced the field by synthesizing pattern-specific code for GPUs, similar to Everest and \name.
While these systems have advanced the state-of-the-art in static graph mining, they do not address the unique challenges posed by temporal constraints in motif mining.

\noindent
\textbf{Temporal Motif Mining}
~introduces additional complexity by incorporating temporal ordering constraints within a specified time window.
Prior work in this field falls into two categories:
(1) \textit{exact methods} like Mackey's chronological edge-matching~\cite{mackey2018chronological} and Everest's GPU acceleration~\cite{everest}, which we extend through co-mining;
(2) \textit{approximate techniques}  estimate the number of matches with high accuracy, significantly reducing computation time from days to minutes~\cite{teacups}. 
They estimate the number of matches by either sampling a subset of edges~\cite{edgesampling}, paths~\cite{teacups} or time intervals ~\cite{sarpe2021presto,liu2019sampling,oden21}.
Oden~\cite{oden21} can estimate multiple motifs with the same underlying structure.
Despite these advancements, most existing systems are limited to CPU-based implementations~\cite{scalable22icde,teacups}, are optimized for only a few motifs~\cite{scalable22icde} or trade accuracy for performance~\cite{sarpe2021presto,liu2019sampling,oden21,edgesampling,oden21}.
\name~provides a flexible framework capable of handling a wide range of motifs, co-mining them efficiently across CPUs and GPUs, without compromising accuracy.

\textbf{Hierarchical Indices:}
FERRARI~\cite{wang2020ferrari} and PRAGUE~\cite{jin2012prague} also exploit hierarchical structures like the MG-Tree, but their goals and designs differ fundamentally from \name.
Both FERRARI’s ADVISE and PRAGUE’s SPIG indexes are built on-the-fly during visual query formulation to record matches of fragments of a single evolving query, and to guide incremental similarity search.
In contrast, the MG-Tree is an offline, compile-time hierarchy over multiple distinct temporal motifs submitted together, grouping them by common edge prefixes (both structural and temporal) to share search paths across all motifs.
While ADVISE and SPIG could accomplish the task of the MG-Tree by building the index with static subgraphs of the motifs and then filtering them for temporal constraints, the need for ADVISE and SPIG to enuemrate their candidates for each fragment of the evolving query results in a large memory and computational overhead.
These overheads are in addition to the computational overhead of enumerating structurally compliant matches before filtering them for temporal constraints~(\S2.1).
The MG-Tree does not suffer from these computational and memory overheads as it does not enumerate all partial matches, and expands them only when temporal and structural constraints are met.
\section{Conclusion}

\name~addresses the critical challenge of efficiently mining multiple temporal motifs by introducing a novel co-mining paradigm that exploits structural and temporal similarities across query patterns.
Our framework introduces the Motif-Group Tree (MG-Tree), a hierarchical data structure that systematically organizes motifs. 
Experimental results demonstrate significant performance improvements, with overall speedup of 2.5$\times$ on the CPU and 1.7$\times$ on the GPU across diverse datasets.
The effectiveness of our approach hinges on the MG-tree exploiting motif similarity to reduce redundant work and exploiting parallelism in the workload.
Architectural bottlenecks posed by enabling co-mining were mitigated by optimizing code-generation.
These advancements not only enhance temporal graph analytics but also underscore the importance of hardware-aware co-design for scalable motif mining.


\bibliographystyle{ACM-Reference-Format}
\bibliography{sample}

\appendix
\section{Design}

\subsection{Definition of MG-Tree}
For a group of temporal motifs $MG = \{M_1, M_2, \ldots, M_k\}$, the MG-Tree $MGT$ is defined as a hierarchical tree of Nodes that capture the similarities among motifs in $MG$, rooted at $N_{root}$.
\begin{description}
    \item[Node Composition:] Any Node $N \in MGT$ is composed of the following 3 members: $C_N$, Children($N$) and $Q_N$,
    \[
    N = \left\langle C_N, \mathrm{Children}(N), Q_N \right\rangle~\text{,where,}
    \]
    \begin{description}
        \item[$C_N$~:-] The (common) motif with edges common across all descendants $C_{N_\text{desc}}$, effectively becoming a prefix for $C_{N_\text{desc}}$ with their first $|C_N|$ being equivalent to $C_N$.
        \[
        \text{prefix}(C_{N_\text{desc}},|C_N|) = C_N
        \]
        
        \item[\(\mathrm{Children}(N)\)~:-] The set of immediate descendants ($N_\text{child}$) that are constructed by directly extending $C_N$, where $C_N$ is the longest prefix of $C_{N_\text{child}}$, not including $C_{N_\text{child}}$.
        \[
        C_N \prec_{\text{max}} C_{N_\text{child}}
        \]
        
        \item[$Q_N$\textbf{:-}] The reference to a query motif $M_i \in MG$ when $C_N$ is equivalent to $M_i$, else is $\emptyset$. For all motifs $M_I \in MG$, there is exactly one Node $N$ in $MGT$ which is responsible for mining $M_i$, i.e. $Q_N = M_i$. Consequently, the union of all nonempty query motif references in $MGT$ is the motif group,
        \[
        MG = \bigcup Q_N ~\forall N\in MGT ~|~ Q_N \neq \emptyset
        \]
    \end{description}
    \item[Root Node:-] $N_{root} \in MGT$ whose common motif $C_{N_{\text{root}}}$ has edges common across all motifs in $MG$.
\end{description}

\subsection{GPU Load-balancing}

Intra-thread Synchronization operations on GPUs are relatively high-lantency operations~\cite{cudaHandbook}, which makes it expensive to monitor threads and calculate a global or even a local balance factor.
Instead, \name simplifies this operation by monitoring the status of threads within a warp periodically (say every INTRA\_INTRVL iterations).
At the end of a period, all threads within a warp vote to perform intra-warp load-balancing if any of the threads are idle and some have work to share.
Monitoring the status of all warps across the GPU can only be done through the global memory, as it is the only piece of memory accessible to all threads across the GPU.
To this end, a globally accessible byte of memory is set when a warp is entirely idle.
Similar to intra-warp balancing, inter-thread balancing is gate-keeped to be performed at a certain interval.
However, as global memory accesses are a lot more expensive that warp-level synchronization~\cite{cudaHandbook}, the period to monitor warps for idleness is a lot longer (INTER\_INTRVL $>$ INTRA\_INTRVL).
Each thread then triggers inter-warp load-balancing when the globally accessible byte of memory is set to indicate idleness.
The pseudocode in Listing ~\ref{lst:gpu-load-balancing} captures the logic of the two-tier load-balancing.

\begin{figure}[h]
\centering
\lstinputlisting[language=Python, caption=Psedocode for GPU-Load Balancing, label=lst:gpu-load-balancing, style=mystyle]{figures/gpu\_load\_balancing.py}
\end{figure}

\subsection{Code-Generation}
Listing~\ref{lst:cpuCode} illustrates the C++ code generated to mine the MG-Tree in  Fig.~\ref{fig:co_mining_search_tree}.

\begin{figure}[]
\begin{lstlisting}[style=mystyle, language=C++, caption=C++ code generated to mine MG-Tree in Fig.~\ref{fig:co_mining_search_tree}., label=lst:cpuCode]
for (Edge e1 : G) { // 1st edge in I1
  list<Edge> cand2_I1 = ...;
  for (Edge e2 : cand2_I1) { // 2nd edge in I1
    list<Edge> cand3_M3 = ...;
    for (Edge e3 : cand3) // 3rd edge in M3
      if (M3.matches({e1, e2, e3, e4}))
        ++count_M3;
    list<Edge> cand3_I2 = ...;
    for (Edge e3 : cand3_I2) { // 3rd edge in I2
      list<Edge> cand4_M4 = ...;
      for (Edge e4 : cand4_M4) // 4th Edge in M4
        if (M4.matches({e1, e2, e3, e4}))
          ++count_M4;
      list<Edge> cand4_M5 = ...;
      for (Edge e4 : cand4_M5) // 4th Edge in M5
        if (M5.matches({e1, e2, e3, e4}))
          ++count_M5;
    }
  }
}
\end{lstlisting}
\end{figure}

\section{Evaluation Results}



%
\subsection{Effect of $\delta$}

To better understand the effect of the size of the temporal window size,$\delta$, on the performance, we expanded this evaluation by,

\begin{enumerate}
    \item Adding four additional $\delta$ configurations: $\delta/4$, $\delta/3$, $3\delta$, and $4\delta$.
    \item Incorporating two other datasets, Stack Overflow (sxo) and Temporal-Reddit-Reply (trr), alongside the original Wikitalk (wtt) dataset.
\end{enumerate}

\begin{figure}[h]
    \begin{subfigure}{\linewidth}
        \centering
        \includegraphics[width=\textwidth]{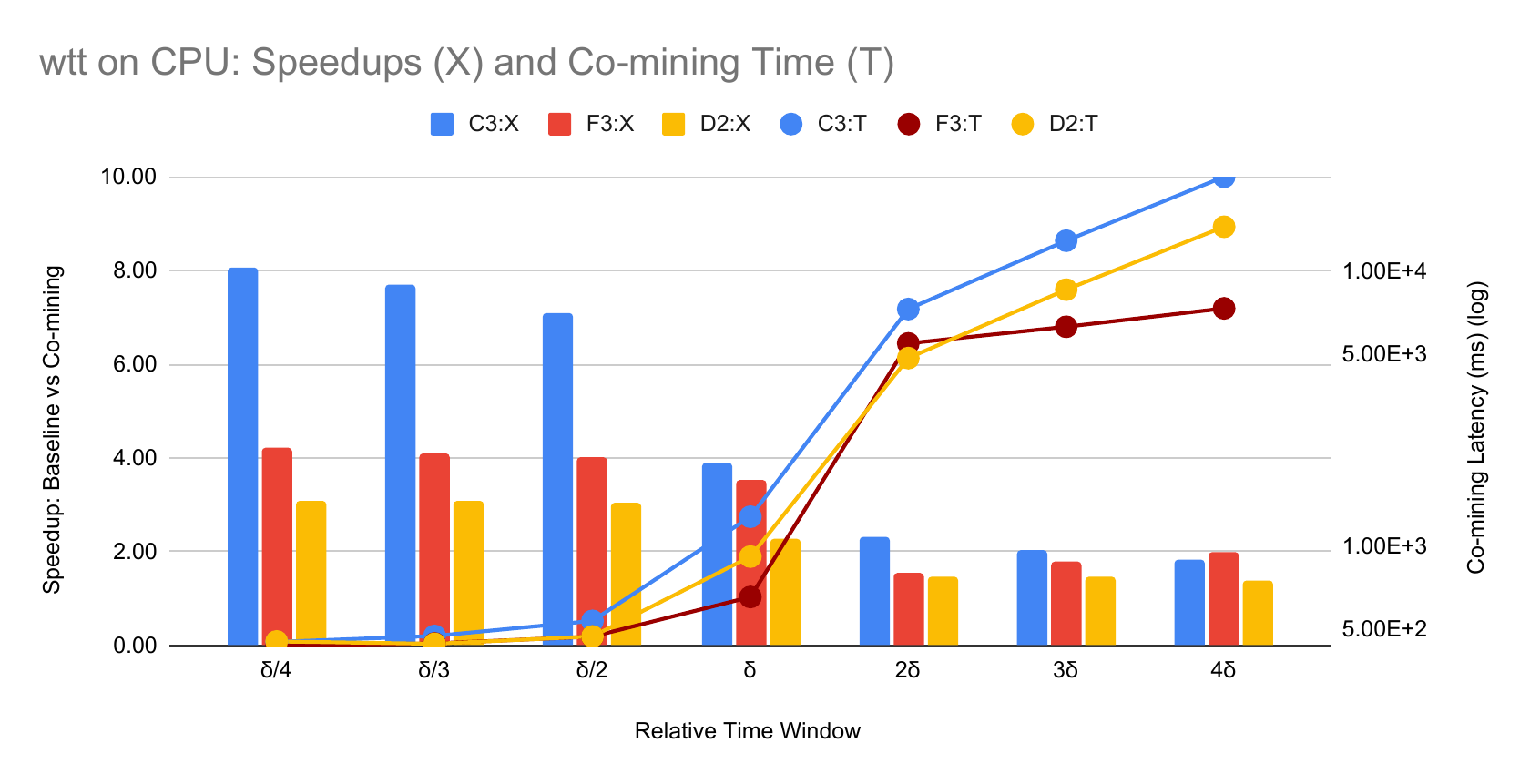}
        \caption{CPU Backend}
        \label{fig:wtt-cpu-delta-ablation-study}
    \end{subfigure}
    \begin{subfigure}{\linewidth}
        \centering
        \includegraphics[width=\textwidth]{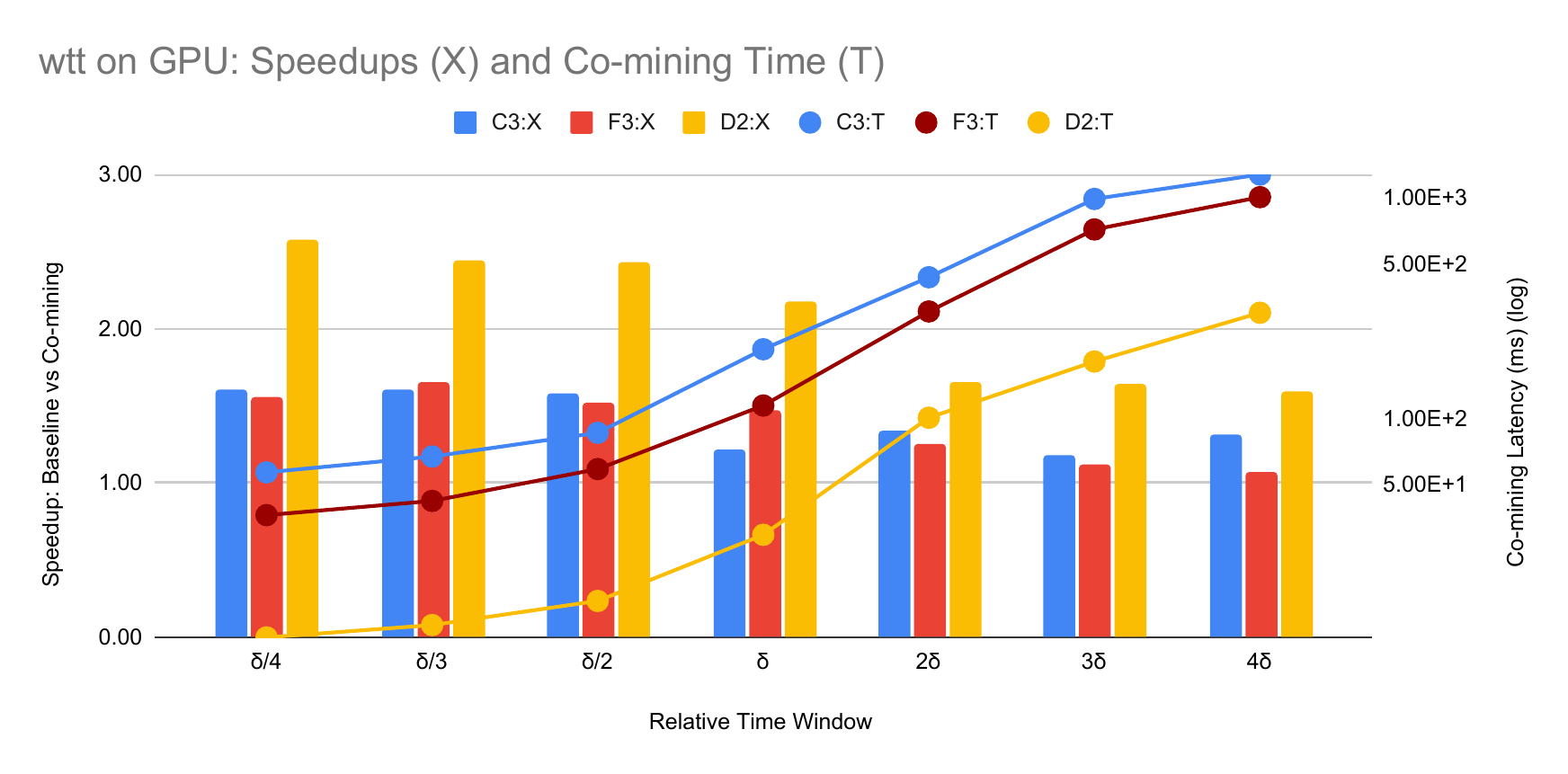}
        \caption{GPU Backend}
        \label{fig:wtt-gpu-delta-ablation-study}
    \end{subfigure}
    \caption{Effect of scaling $\delta$ on Speedup and Runtime on the Wikitalk (wtt) dataset.}
    \label{fig:wtt-delta-study}
\end{figure}

\begin{figure}[h]
    \begin{subfigure}{\linewidth}
        \centering
        \includegraphics[width=\textwidth]{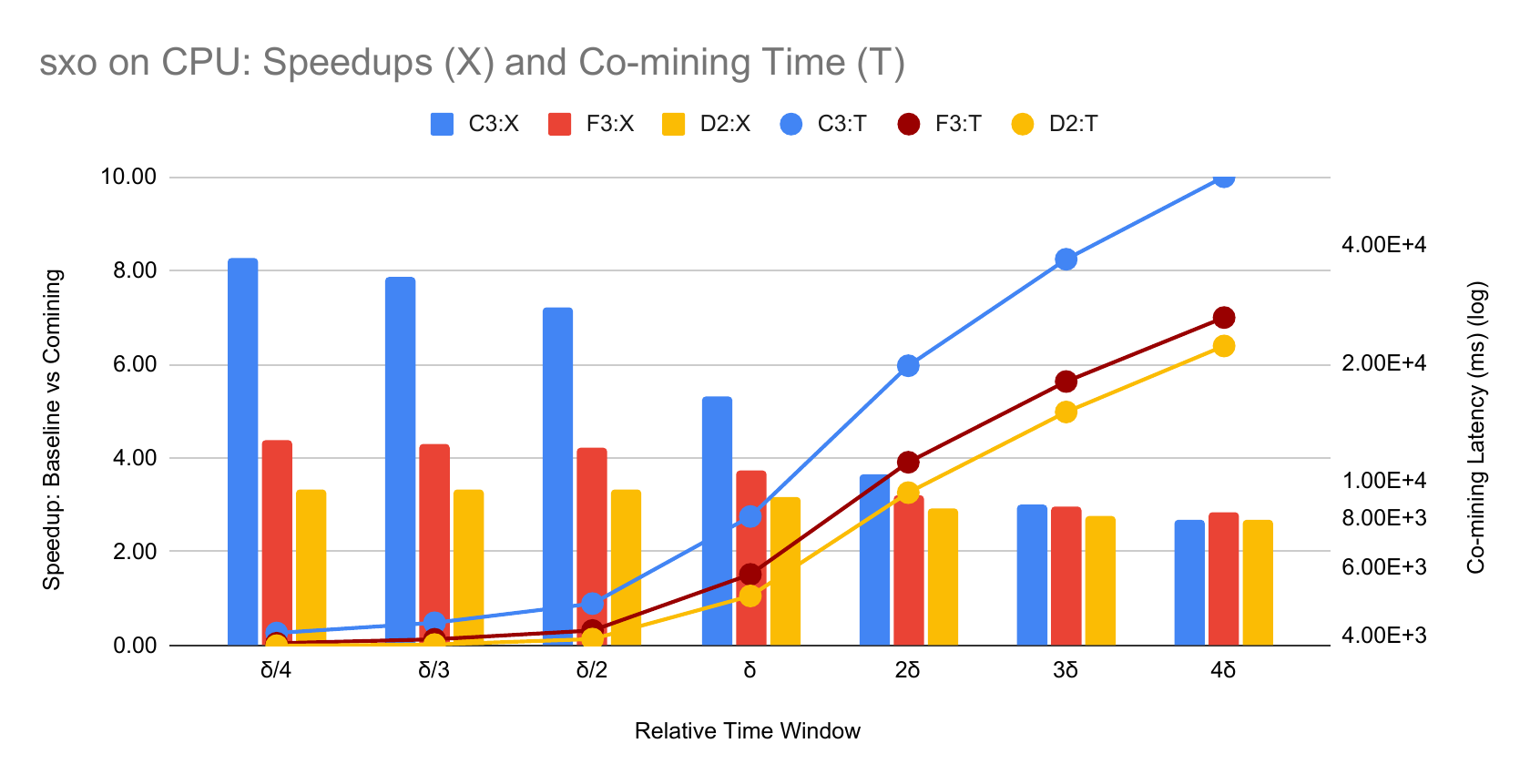}
        \caption{CPU Backend}
        \label{fig:eqx-cpu-delta-ablation-study}
    \end{subfigure}
    \begin{subfigure}{\linewidth}
        \centering
        \includegraphics[width=\textwidth]{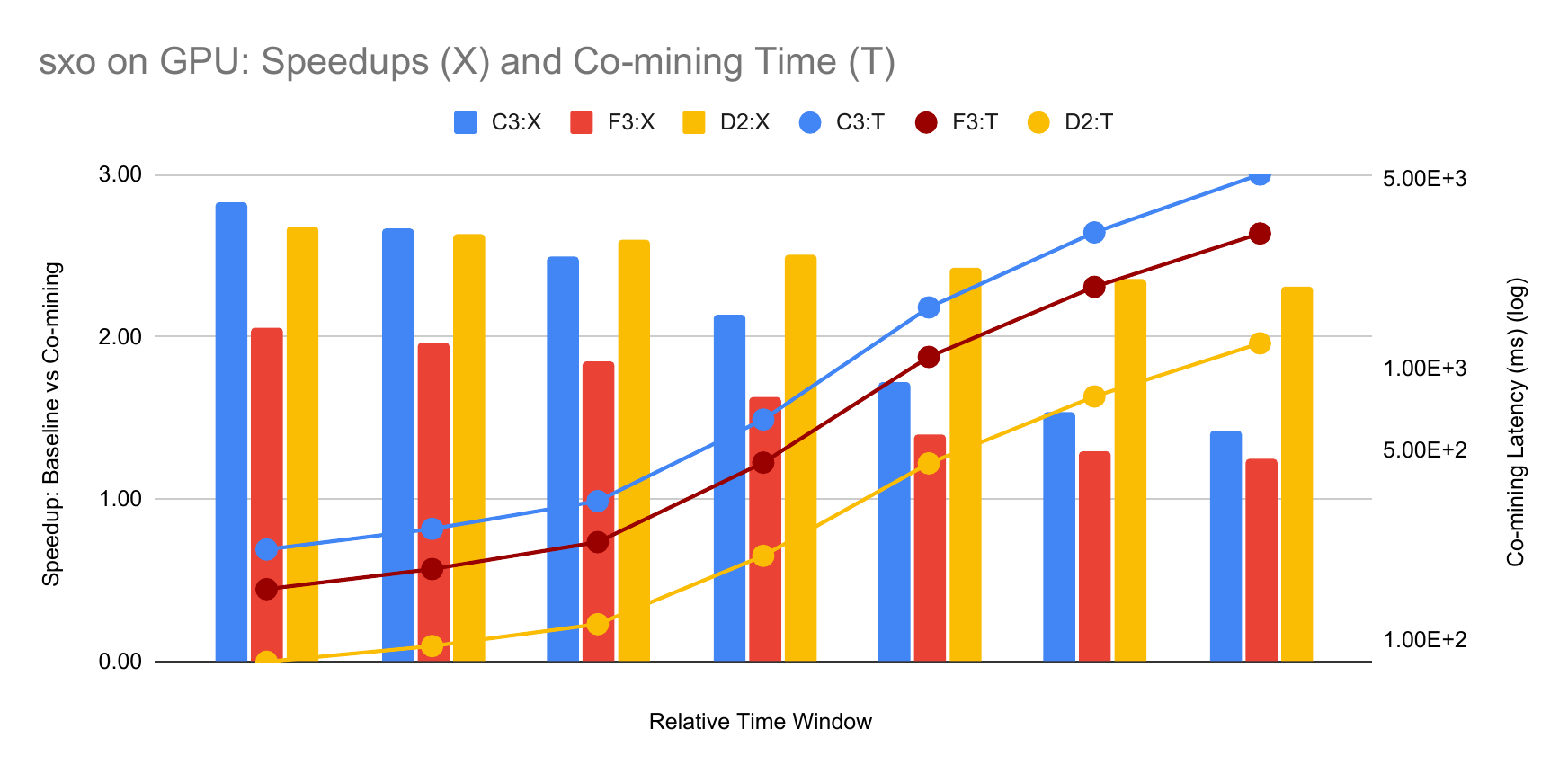}
        \caption{GPU Backend}
        \label{fig:eqx-gpu-delta-ablation-study}
    \end{subfigure}
    \caption{Effect of scaling $\delta$ on Speedup and Runtime on the Stack Overflow (sxo) dataset.}
    \label{fig:sxo-delta-study}
\end{figure}

\begin{figure}[h]
    \begin{subfigure}{\linewidth}
        \centering
        \includegraphics[width=\textwidth]{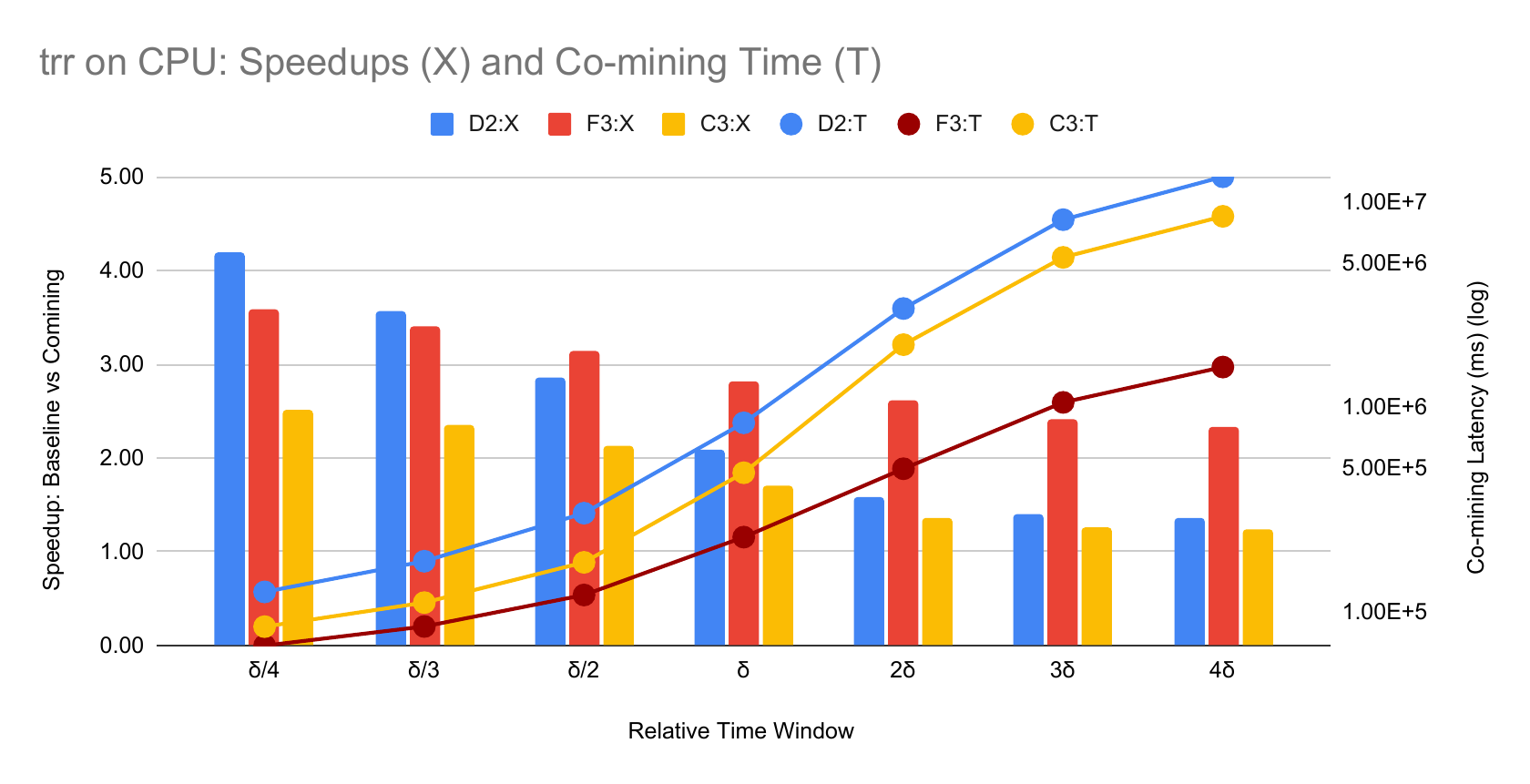}
        \caption{CPU Backend}
        \label{fig:trr-cpu-delta-ablation-study}
    \end{subfigure}
    \begin{subfigure}{\linewidth}
        \centering
        \includegraphics[width=\textwidth]{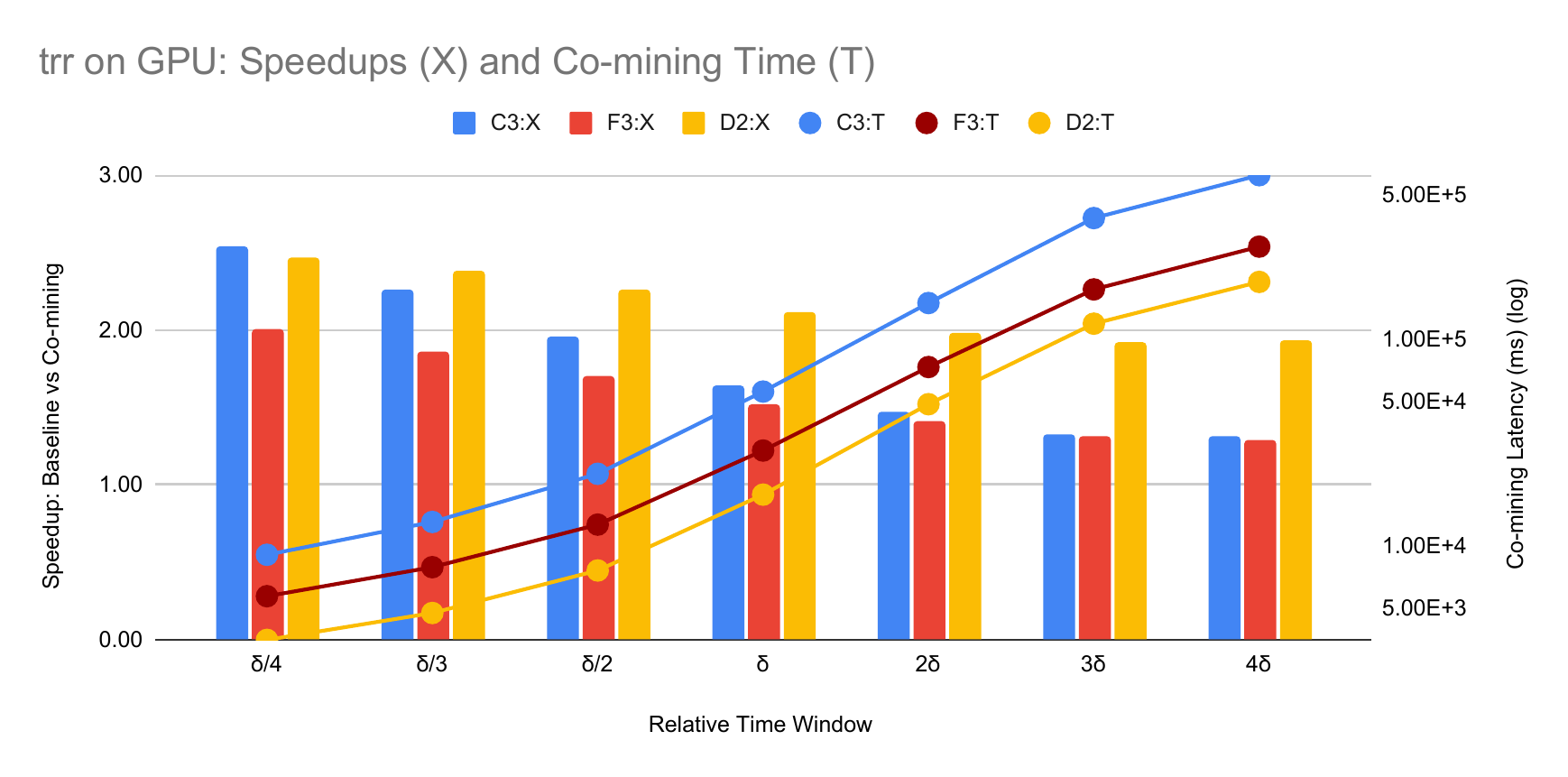}
        \caption{GPU Backend}
        \label{fig:trr-gpu-delta-ablation-study}
    \end{subfigure}
    \caption{Effect of scaling $\delta$ on Speedup and Runtime on the trr dataset.}
    \label{fig:trr-delta-study}
\end{figure}

Our expanded results (Figures~\ref{fig:wtt-delta-study},~\ref{fig:sxo-delta-study},~\ref{fig:trr-delta-study}) reinforces the observations from the smaller scale experiment in the paper: Increasing the time window ($\delta$) diminishes the performance advantage of co-mining over the baseline.
This is because larger window sizes expand the candidate search space across all levels of the MG-Tree hierarchy, and sequentializing the exploration until the system can re-balance the load.
While the GPU backend is able to load-balance across all levels of the MG-Tree, the CPU backend is only able to do so at the top-most level, leading to a larger loss in the performance gap between the baseline and the co-mining execution.
This is evident from the ration between the speedup achieved at $\delta/4$ and $4\delta$ in Table~\ref{tab:speedup-ratio}.

\begin{table}[h]
\begin{tabular}{c}
            \begin{tabular}{|r|r|r|}
            \hline
            \multicolumn{1}{|l|}{\textbf{MG}} & \multicolumn{1}{l|}{\textbf{GPU Ratio}} & \multicolumn{1}{l|}{\textbf{CPU Ratio}} \\ \hline
            D2 & 1.62 & 2.22 \\ \hline
            F3 & 1.54 & 2.71 \\ \hline
            C3 & 1.36 & 4.42 \\ \hline
            \end{tabular}\\
            Wikitalk (wtt)\\
    
            \begin{tabular}{|r|r|r|}
            \hline
            \multicolumn{1}{|l|}{\textbf{MG}} & \multicolumn{1}{l|}{\textbf{GPU Ratio}} & \multicolumn{1}{l|}{\textbf{CPU Ratio}} \\ \hline
            D2 & 1.62 & 2.22 \\ \hline
            F3 & 1.54 & 2.71 \\ \hline
            C3 & 1.36 & 4.42 \\ \hline
            \end{tabular}\\
            Stack-Overflow (sxo)\\

            \begin{tabular}{|r|r|r|}
            \hline
            \multicolumn{1}{|l|}{\textbf{MG}} & \multicolumn{1}{l|}{\textbf{GPU Ratio}} & \multicolumn{1}{l|}{\textbf{CPU Ratio}} \\ \hline
            D2 & 1.62 & 2.22 \\ \hline
            F3 & 1.54 & 2.71 \\ \hline
            C3 & 1.36 & 4.42 \\ \hline
            \end{tabular}\\
            Temporal Reddit Reply (trr)\\
    
\end{tabular}
\caption{Ratio between the speedup at $\delta/4$ vs $4\delta$ for different datasets.}
\label{tab:speedup-ratio}
\end{table}

\end{document}